\documentclass[twocolumn, showpacs]{revtex4}
\usepackage{graphicx}
\usepackage{epstopdf}

\usepackage{color}
\graphicspath{{pix/}}

\begin{document}

\title{Comparative study of nonclassicality, entanglement, and dimensionality of multimode noisy twin beams}

\author{Ievgen I. Arkhipov}
\email{arkhipov@jointlab.upol.cz}
\address{RCPTM, Joint Laboratory of Optics of Palack\'y University and
Institute of Physics of AS CR, Palack\'y University, 17. listopadu
12, 771 46 Olomouc, Czech Republic}

\author{Jan Pe\v{r}ina Jr.}
\address{RCPTM, Joint Laboratory of Optics of Palack\'y University and
Institute of Physics of AS CR, Palack\'y University, 17. listopadu
12, 771 46 Olomouc, Czech Republic}

\author{Adam Miranowicz}
\address{Faculty of Physics, Adam Mickiewicz University, PL-61-614 Poznan, Poland}

\author{Jan Pe\v{r}ina}
\address{RCPTM, Joint Laboratory of Optics of Palack\'y University and
Institute of Physics of AS CR, Palack\'y University, 17. listopadu
12, 771 46 Olomouc, Czech Republic}

\begin{abstract}
The nonclassicality, entanglement, and dimensionality of a
noisy twin beam are determined using a characteristic function of
the beam written in the Fock basis. One-to-one correspondence
between the negativity quantifying entanglement and the
nonclassicality depth is revealed. Twin beams, which are either
entangled or nonclassical (independent of their entanglement), are
observed only for the limited degrees of noise, which degrades their
quantumness. The dimensionality of the twin beam quantified by the
participation ratio is compared with the dimensionality of
entanglement determined from the negativity. Partitioning of the
degrees of freedom of the twin beam into those related to
entanglement and to noise is suggested. Both single-mode and
multimode twin beams are analyzed. Weak nonclassicality based on
integrated-intensity quasidistributions of multimode twin beams is
studied. The relation of the model to the experimental twin beams is
discussed.
\end{abstract}


\maketitle

\section{Introduction}

The question whether a given state cannot be described within a
classical theory has been considered  one of the most serious
since the early days of quantum
physics~\cite{Einstein05,Einstein35,Schrodinger35} (for a review
see, e.g., Ref.~\cite{Perina1994Book}). Nonclassicality and
entanglement, which is one of the nonclassicality manifestations,
are the most important properties of optical fields studied in
quantum optics. Such fields have no classical analogs and as such
they have been found interesting for many reasons. Nonclassical
properties of such fields have been found useful both for
elucidating the principles of quantum mechanics and in
various applications including, e.g., quantum information
processing~\cite{NielsenBook}, quantum
metrology~\cite{Migdall1999,Giovannetti2011,PerinaJr2012a}, and
highly-sensitive measurements~\cite{Vahlbruch10}.

From both the theoretical and the experimental points of view, the
nonlinear process of parametric down-conversion, in which photon
pairs are generated, has played an important role here from the
beginning of investigations~\cite{MandelBook,
Perina1991Book,Perina1994Book,WallsBook}. Its individual photon
pairs have been exploited in many fundamental experiments testing
nonclassical behavior predicted by quantum
physics~\cite{Bouwmeester1997,Weihs1998}. It has also allowed the
generation of more intense fields having their electric-field amplitude
quadratures squeezed below the vacuum
level~\cite{Luks1988,Lee1990,Lvovsky2009}, exhibiting
sub-shot-noise correlations~\cite{Jedrkiewicz2004,Bondani2007} or
having sub-Poissonian photon-number
statistics~\cite{PerinaJr2013b,Lamperti2014,Lamperti2014a}.

In quantum optics, the definition of nonclassicality is based upon
the Glauber-Sudarshan $P$
representation~\cite{Glauber63,Sudarshan63,Perina1991Book} of the
statistical operator of a given field. The commonly accepted
formal criterion for distinguishing nonclassical states from
classical ones is expressed as
follows~\cite{DodonovBook,VogelBook,MandelBook,Perina1991Book}: A
quantum state is {\it nonclassical} if its Glauber-Sudarshan $P$
function fails to have the properties of a probability density.
Alternatively, several operational criteria for nonclassicality of
either single-mode~\cite{DodonovBook,VogelBook,Richter02} or
multimode~\cite{Miran10,Bartkowiak11,Allevi2013} fields have been
revealed. Their derivations are based either on field's
moments~\cite{Vogel08,Miran10,Allevi2013} or on direct reconstruction
of quasidistributions of integrated
intensities~\cite{Haderka2005a,PerinaJr2012,PerinaJr2013a}. Also,
criteria derived from the majorization theory have been
found~\cite{Lee1990a,Verma2010}.

Entanglement (or inseparability) is a special nonclassical
property that describes quantum correlations among (in general)
several subsystems that cannot be treated by the means of
classical statistical theory~\cite{Horodecki09review}. Various
approaches have been developed for discrete and continuous
variables to reveal entanglement. This property has been exploited
in suggesting an entanglement criterion and the related
entanglement measure (referred to as the negativity) based upon
the partial transposition of a statistical
operator~\cite{Peres96,Horodecki97,Zyczkowski98,Vidal02}. Another
approach has been based on the violation of the Bell inequalities
written for different mean values including the measurement on
both parts of a bipartite system~\cite{CHSH}. Also, a method using
positive semi-definite matrices of fields' moments of different
orders~\cite{Shchukin05,Miran09} has been found to be very powerful. We
would like to stress at this point that entanglement is a very
crucial tool in today's quantum information processing.

In this contribution, we study nonclassicality by applying the
Lee nonclassical depth~\cite{Lee91} as well as entanglement via
the negativity~\cite{Zyczkowski98,Vidal02} for (in general) noisy
twin beams of different intensities. Such fields occur under real
experimental conditions in which a nonlinear crystal generates
both photon pairs and individual single photons (noise).
Nevertheless, the signal and idler fields together form a
bipartite quantum system. We note that entanglement and
nonclassicality of twin beams generated by down-conversion seeded
by thermal light have been analyzed in
Refs.~\cite{Degiovanni2007,Bondani2008,Degiovanni2009}. In this
case, noise present in the incident thermal fields participates in
the nonlinear process and generation of photon pairs. This weakens
its detrimental effect on entanglement and nonclassicality of twin
beams and allows us to have entangled twin beams with a larger amount
of noise.

Here we also study the problem of entanglement dimension via the
negativity $N$ for general twin beams and the Schmidt number $K$
for noiseless twin beams in a pure state. Namely, we estimate how
many degrees of freedom of two fields comprising a twin beam are
entangled based on the results of Ref.~\cite{Eltschka13} for
axisymmetric states. On the other hand, the participation ratio $
R_{\rm s} $~\cite{Zyczkowski99} determined from the reduced
statistical operator $\hat\rho_{\rm s}$ of the signal (or idler)
field gives the number of degrees of freedom in this field serving
to describe both entanglement and noise. It varies from $R_{\rm
s}=1$ (for a pure state $\hat\rho_{\rm s}$) to $R_{\rm
s}=d=\dim(\hat\rho_{\rm s})$ for the completely mixed state
$\hat\rho_{\rm s}=I/d$. We note that the participation ratio $
R_{\rm s} $ gives an effective number of states in the mixture
$\hat\rho_{\rm s}$ implied by the property that it is a lower
bound for the rank of $\hat\rho_{\rm s}$. Moreover, the logarithm
of $R$ is the von Neumann--Renyi entropy of second
order~\cite{Zyczkowski99}. The inverse of the participation ratio
is referred to as the purity (or linear entropy). Various methods
for direct measuring the Schmidt number $ K $ (even without
recourse to quantum tomography) were proposed for noiseless twin
beams (see, e.g.,
Refs.~\cite{Fedorov04,Fedorov07,Chan07,Pires09,Bartkiewicz13}).
The method of Ref.~\cite{Chan07} was recently realized
experimentally~\cite{Just13}. We note that the negativity can also
be measured without applying quantum tomography as described,
e.g., for two polarization qubits using linear optical
setups~\cite{Bartkiewicz14,Bartkiewicz14a}.

The paper is organized as follows. In Sec.~II, the model of
parametric down-conversion providing an appropriate statistical
operator of a twin beam is presented. Entanglement of the twin
beam is addressed in Sec.~III using the negativity. The
nonclassical depth is introduced in Sec.~IV to quantify
nonclassicality. The relation between the negativity and the nonclassical
depth is also discussed in Sec.~IV. The dimensionality of a twin beam
described by the participation ratio together with the entanglement
dimensionality described by the negativity is analyzed in Sec.~V.
Properties of $ M $-mode twin beams are discussed in Sec.~VI.
Section~VII is devoted to experimental multimode twin beams
containing also noise embedded in independent spatiotemporal
modes. Conclusions are drawn in Sec.~VIII.

\section{Quantum model of a twin beam}

To describe the generation of a single-mode twin beam by
parametric down-conversion, we adopt the approach based on the
Heisenberg equations derived from the appropriate nonlinear
Hamiltonian $\hat H_{\rm int}$~\cite{Perina1991Book},
\begin{equation}  
\hat H_{\mathrm{int}} = -\hbar g\hat A_{1}\hat A_{2}\exp(\mathrm
i\omega t-\mathrm i\phi) + \mathrm H.\mathrm c.,
\end{equation}
where $\hat A_{1}$ $(\hat A^{\dagger}_{1})$ and $\hat A_{2}$
$(\hat A^{\dagger}_{2})$ represent the annihilation (creation)
operators of the signal and idler field, respectively, and $g$ is a
real coupling constant that is linearly proportional both to the
quadratic susceptibility of a nonlinear medium and to the real
pump-field amplitude. The interaction time is denoted  $t$, $\omega
$ ($ \phi $) is the pump-field frequency (phase), and
$\omega_{1}$ and $\omega_{2}$ stand for the signal- and
idler-field frequencies, respectively. The law of energy
conservation provides the relation $\omega = \omega_{1} +
\omega_{2}$. H.c. is the Hermitian conjugated term. In a
real nonlinear process, also noise occurs. It can be described by
the Langevin forces $\hat L$ belonging to a reservoir of chaotic
oscillators with mean number of noise photons $\langle n_{\rm
d}\rangle$.

The Heisenberg-Langevin equations corresponding to the Hamiltonian $
\hat H_{\mathrm{int}} $ are written as
\begin{eqnarray} \label{hle} 
 \frac{\mathrm d\hat A_{1}}{\mathrm dt} &=& -(\mathrm i\omega_{1} + \gamma_1 ) \hat A_{1} +
  \mathrm ig\hat A^{\dagger}_{2}\exp(-\mathrm i\omega t+ \mathrm i\phi) + \hat L_{1}, \nonumber \\
 \frac{\mathrm d\hat A_{2}}{\mathrm dt} &=& -(\mathrm i\omega_{2}+ \gamma_2 ) \hat
  A_{2} + \mathrm ig\hat A^{\dagger}_{1}\exp(-\mathrm i\omega t+
  \mathrm i\phi) + \hat L_{2}, \nonumber \\
 & &
\end{eqnarray}
where the constant $\gamma_1 $ ($ \gamma_2 $) describes damping in
the signal (idler) field. The Langevin operators $\hat L_i $ (for
$i = 1,2$) have the properties
\begin{eqnarray}  
& \langle\hat L_{i}\rangle = \langle\hat L^{\dagger}_{i}\rangle=
 0, \langle\hat L^{\dagger}_{i}\hat L_{j}\rangle =
2\gamma_j\langle n_{\rm d}\rangle \delta_{ij}, & \nonumber \\
& \langle\hat L_{i}\hat L^{\dagger}_{j}\rangle =2\gamma_j\left(
\langle n_{\rm d}\rangle +1\right) \delta_{ij}, &
\end{eqnarray}
where $\delta_{ij}$ stands for the Kronecker symbol.

Using the interaction representation [$ \hat A_{j}(t) = a_{j}(t)
\exp(-i\omega_j t) $] and neglecting damping together with the
Langevin forces, the solution of Eq.~(\ref{hle}) attains the form
\begin{eqnarray}    
 \hat a_{1}(t) &=& \hat a_{1}(0)u(t) + i\hat a^{\dagger}_{2}(0)v(t)\exp(i\phi),
  \nonumber \\
 \hat a_{2}(t) &=& \hat a_{2}(0)u(t) + i\hat
   a^{\dagger}_{1}(0)v(t)\exp(i\phi),
\label{4}
\end{eqnarray}
in which $u(t) = \cosh(gt)$ and $v(t) = \sinh(gt)$.

Statistical properties of the twin beam are then described by the
normal characteristic function $ C_{\mathcal N} $ defined as
\begin{equation}    
C_{\mathcal N}(\beta_{1},\beta_{2}) =
\mathrm{Tr}\left[\hat\rho\exp (\beta_{1}\hat
a^{\dagger}_{1}+\beta_{2}\hat
a^{\dagger}_{2})\exp(-\beta^{\ast}_{1}\hat a_{1}
-\beta^{\ast}_{2}\hat a_{2})\right],
\end{equation}
where ${\rm Tr}$ denotes the trace. Using the solution given in
Eq.~(\ref{4}), the normal characteristic function $ C_{\mathcal N}
$ attains the Gaussian form~\cite{Perina05},
\begin{eqnarray}       
 C_{\mathcal N}(\beta_{1},\beta_{2})&=&\exp\left[-
  (\vert\beta_{1}\vert^{2}
  B_{1}+\vert\beta_{2}\vert^{2}B_{2})+D_{12}\beta^{\ast}_{1}\beta^{\ast}_{2}+ \nonumber \right. \nonumber \\
 & & \left. \hspace{8mm} +D^{\ast}_{12}\beta_{1}\beta_{2}\right],
\end{eqnarray}
in which $\beta_{1} $ and $\beta_{2}$ denote independent complex
variables. For the undamped and noiseless case, we have $ D_{12} =
\langle\bigtriangleup\hat a_{1}\bigtriangleup\hat a_{2}\rangle$.
Also the mean number $ B_{\rm p} $ of the generated photon pairs
is determined as $B_{\rm p} = \langle\bigtriangleup\hat
a^{\dagger}_{1}\bigtriangleup\hat a_{1}\rangle=
\langle\bigtriangleup\hat a^{\dagger}_{2}\bigtriangleup\hat
a_{2}\rangle$. When damping and noise are also
considered~\cite{Perina05}, the parameters $ B_{a} $ (for $ a=1,2
$) contain additional noise contributions characterized by the
parameters $ B_{\rm s}$ and $ B_{\rm i}$, i.e., $B_{1} = B_{\rm p}
+ B_{\rm s} $ and $B_{2} = B_{\rm p} + B_{\rm i} $. Whereas the
parameter $ B_{\rm p} $ gives the mean number of photon pairs, the
parameters $ B_{\rm s} $ and $ B_{\rm i} $ correspond to the mean
number of noise photons coming from the signal- and idler-field
reservoirs, respectively. On the other hand, the parameter
$D_{12}$ describing mutual correlations between the signal and
idler fields is not influenced by the noise since $ |D_{12}|^2 =
B_{\rm p}(B_{\rm p}+1) $.

The statistical operator $\hat\rho$ of the twin beam then acquires
the form~\cite{Perina1991Book}
\begin{equation} \label{rho}    
 \vspace{2mm}\hat\rho = \frac{1}{\pi^2} \int d^{2} \beta_{1}
 d^{2}\beta_{2} \mathcal{C}_{\mathcal A}(\beta_{1},\beta_{2})
 :\exp \left(\sum_{j=1}^{2} \hat a_{j}\beta_{j}^{\ast}-\hat
 a_{j}^{\dagger}\beta_{j}\right):.
\end{equation}
In Eq.~(\ref{rho}), $C_{\mathcal A}(\beta_{1},\beta_{2}) =
C_{\mathcal
N}(\beta_{1},\beta_{2})\exp(-\vert\beta_{1}\vert^{2}-\vert\beta_{2}\vert^{2})$
denotes an anti-normal characteristic function and symbol $ :$  $:$
means normal ordering of field operators.

Performing integration in Eq.~(\ref{rho}) we express the
statistical operator $ \hat\rho $ in the form
\begin{eqnarray}   \label{rhofor}  
 \hat\rho &=& \frac{1}{\tilde K} : \exp\left[ -\frac{\tilde B_{2}}{\tilde K}\hat a_{1}^\dagger\hat a_{1}
  - \frac{\tilde B_{1}}{\tilde K}\hat a_{2}^\dagger\hat a_{2} \nonumber \right.\\
 & & \left.\hspace{14mm}+\frac{|D_{12}|}{\tilde K}
  \left(\hat a_{1}\hat a_{2} + \hat a_{1}^{\dagger}\hat
  a_{2}^{\dagger}\right)\right]:,
\end{eqnarray}
where $\tilde K = \tilde B_{1}\tilde B_{2}-\vert D_{12}\vert^{2}$.
The parameters $ \tilde B_{a}$ introduced in Eq.~(\ref{rhofor})
are related to anti-normal ordering of field operators and are
given as $ \tilde B_{a} = B_{a} + 1$ with $a=1,2 $. Decomposing
the statistical operator $ \hat\rho $ in the Fock-state basis we
finally arrive at the formula
\begin{eqnarray}\label{frho}   
 \rho_{ij,kl} &=& \sum_{n=0}^{\infty}\sum_{p=0}^{n}\sum_{r=0}^{p}\sum_{t=0}^{r}
  (-1)^{n-r}\frac{\tilde{B}^{n-p}_{2}\tilde{B}^{p-r}_{1}\tilde{K}^{-n-1}}{(n-p)!(p-r)!} \nonumber \\
  & & \hspace{-8mm} \times\frac{|D_{12}|^{r}}{(r-t)!\, t!}
   \langle ij|\hat a_{1}^{\dagger n-p+t}\hat a_{2}^{\dagger p-r+t}
   \hat a_{1}^{n-p+r-t}\hat a_{2}^{p-t}| kl\rangle. \nonumber \\
  & &
\end{eqnarray}

Direct inspection of Eq.~(\ref{frho}) for the matrix elements of the
statistical operator $\hat\rho$ written in Eq.~(\ref{frho})
reveals that all nonzero elements can be parameterized by only
three indices,
\begin{eqnarray}\label{rho1}   
 \rho_{i,j,i+d,j+d} &=& \frac{1}{\tilde
   K}\sqrt{\frac{(i+d)!}{i!}\frac{(j+d)!}{j!}}\sum\limits_{m=0}^{\mathrm{max}(i,j)}
   C^i_m C^j_m \nonumber \\
 & & \hspace{-8mm} \times
  \frac{m!}{(m+d)!}X_1^{j-m}
 X_2^{i-m}\left(\frac{\vert D_{12}\vert}{\tilde
  K}\right)^{d+2m},
\end{eqnarray}
assuming $d \geq 0$. Moreover, $\rho_{ij,i+d,j+d}=\rho_{i+d,j+d,
i,j}$, $X_a=1-\tilde B_{a}/\tilde K$ with $a=1,2$, and $C^i_m$ and
$C^j_m$ denote the binomial coefficients.

\section{Negativity of the twin beam}

The negativity $ N $ of a mixed bipartite system defined on the
basis of the Peres-Horodecki criterion for a partially transposed
statistical operator~\cite{Peres96,Horodecki97,Vidal02} is useful
for quantifying the entanglement of the twin beam. It can be expressed
as
\begin{equation} \label{neg}    
 N(\hat\rho) = \frac{\vert\vert\hat\rho^{{\Gamma}}\vert\vert_{1}-1}{2}
\end{equation}
using the trace norm $\vert\vert\rho^{{\Gamma}}\vert\vert_{1}$ of
the partially transposed statistical operator $ \rho^{{\Gamma}} $.
The negativity essentially measures the degree at which
$\rho^{\Gamma}$ fails to be positive. As such it can be regarded
as a quantitative version of the Peres-Horodecki criterion for
separability~\cite{Peres96,Horodecki97}. According to
Eq.~(\ref{neg}), the negativity $ N $ is given as the absolute
value of the sum of the negative eigenvalues of $\rho^{\Gamma}$.
It vanishes for separable states. It is worth noting that the
negativity $ N$ is an entanglement monotone and so it can be used
to quantify the degree of entanglement in bipartite systems.
Moreover, the negativity does not reveal bound entanglement (i.e.,
nondistillable entanglement) in systems more complicated than two
qubits or qubit-qutrit~\cite{Horodecki09review}.

To determine the negativity $ N $ we consider the eigenvalue
problem for the partially transposed statistical operator
$\hat\rho^{\Gamma} $. The statistical operator $\hat\rho^{\Gamma} $
expressed in the Fock-state basis attains a characteristic block
structure. The smallest block has dimension 2 and each
successive block has dimension larger by 1. For a  given $M$ one
has a block of dimension $M+1$. Such a block represents a
matrix of $M+1$ isolated states; the sum of indices of their
statistical operators equals $2M$,
\begin{equation}\label{matrix}    
\hat\rho^{\Gamma}_M=\left( \begin{array}{cccc}
  \rho_{0\,M,0\,M} & \rho_{0\,M-1,1\,M} &\dots &\rho_{0\,0,M\,M} \\
  \rho_{1\,M,0\,M-1} & \rho_{1\,M-1,1\,M-1} &\dots &\dots \\
  \dots & \dots & \dots &\dots \\
 \rho_{M\,M,0\,0} & \dots &\dots &\rho_{M\,0,M\,0} \\
   \end{array} \right).
\end{equation}

It can be shown that eigenvalues of a block of dimension $M+1$ can
be expressed as $\nu_{+}^M, \nu_{+}^{M-1}\nu_{-},\dots ,
\nu_{+}\nu_{-}^{M-1},\nu_{-}^{M}$ using the eigenvalues $ \nu_+ $
and $ \nu_- $ of a block with dimension 2:
\begin{equation}               
\nu_{\pm} = 1-\frac{1}{2\tilde K}\left(\tilde B_{1}+\tilde
B_{2}\mp\sqrt{\left(\tilde B_{2}-\tilde B_{1}\right)^2+4\vert
D_{12}\vert^2}\right).
\end{equation}
The negative eigenvalues can only be those containing odd powers
of $\nu_{-}$. They form a geometric progression whose elements can
be summed to arrive at the formula for the negativity $ N $:
\begin{equation}\label{fneg}       
 N =\frac{1}{2} \frac{3(\tilde B_{1}+\tilde B_{2})+\sqrt{(\tilde B_{1}-\tilde B_{2})^2+4\vert
  D_{12}\vert^2}-4\tilde K-2}{4\tilde K-2(\tilde B_{1}+\tilde
  B_{2})+1}.
\end{equation}
Expressing parameters $ \tilde B_1 $, $ \tilde B_2 $, and $
|D_{12}|^2 $ in Eq.~(\ref{fneg}) in terms of parameters $ B_{\rm
p} $, $ B_{\rm s} $, and $ B_{\rm i} $, we arrive at the formula
\begin{eqnarray}\label{fneg1}       
 N &=& \bigl\{ 2B_{\rm p}-(B_{\rm s}+B_{\rm i})(4B_{\rm p}+1)-4B_{\rm s}B_{\rm i}\nonumber \\
 & & \mbox{} +\sqrt{(B_{\rm s}-B_{\rm i})^2+4B_{\rm p}(B_{\rm p}+1)} \bigr\} \nonumber \\
 & & \mbox{} \times \bigl\{
 4(B_{\rm s}+B_{\rm i})(2B_{\rm p}+1)+8B_{\rm s}B_{\rm i}+2\bigr\}^{-1}.
\end{eqnarray}
Equation~(\ref{fneg1}) simplifies considerably  for noiseless twin
beams:
\begin{equation}\label{negn}    
 N = B_{\rm p} + \sqrt{B_{\rm p}(B_{\rm p}+1)}.
\end{equation}

\begin{figure}  
 \includegraphics[width=.48\textwidth]{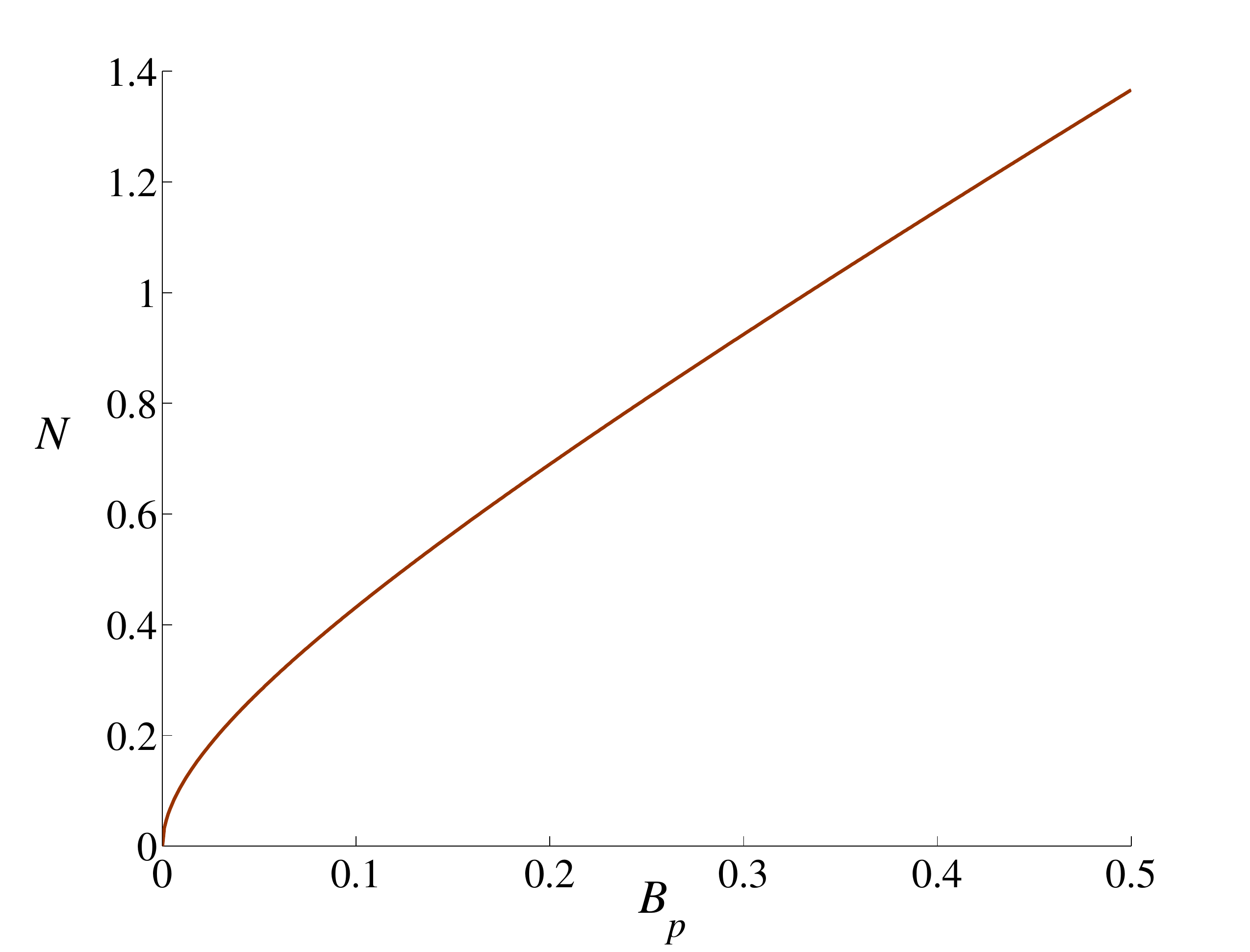}
 \caption{(Color online) Negativity $ N $ as a function of the mean photon-pair
  number $ B_{\rm p} $ for noiseless twin beams (i.e., $ B_{\rm s} = B_{\rm i} = 0)$)
  according to Eq.~(\ref{negn}).}
\label{fig1}
\end{figure}

\begin{figure}          
 \includegraphics[width=0.48\textwidth]{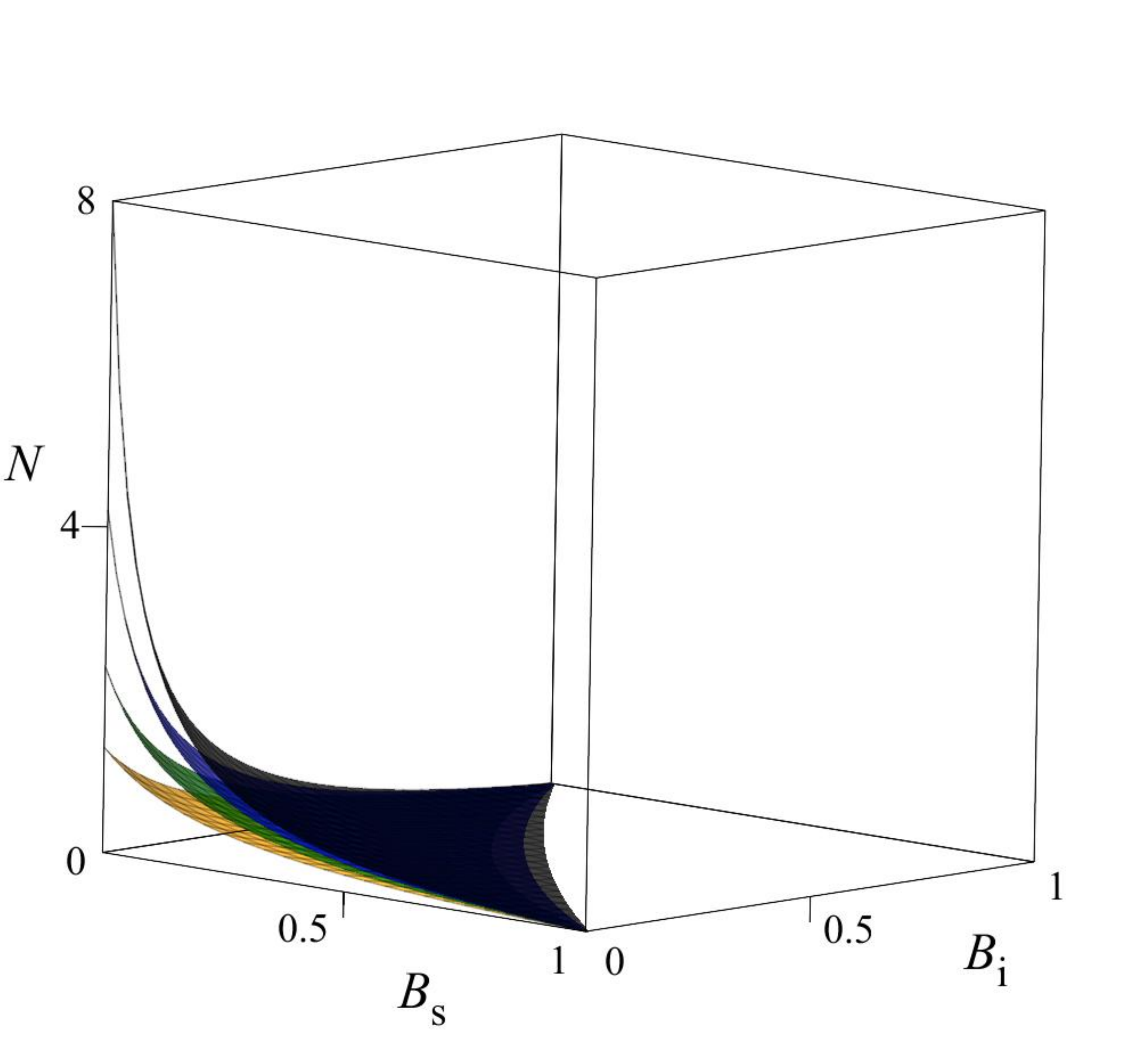}
 \caption{(Color online)
  Negativity $ N $, given in Eq.~(\ref{fneg1}), as a function of the mean noise
  photon numbers $B_{\rm s}$ and $B_{\rm i}$ in the signal and
  idler modes, respectively, assuming the mean photon-pair
  number $B_{\rm p}$ equal to $0.5$  [bottom light-gray (yellow) area], $1$ [gray (green) area],
  $2$ [dark-gray (blue) area] and $4$ [top, black area].
  The larger $ B_{\rm p} $, the larger the negativity $ N $.}
\label{fig2}
\end{figure}

\begin{figure}          
 \includegraphics[width=0.48\textwidth]{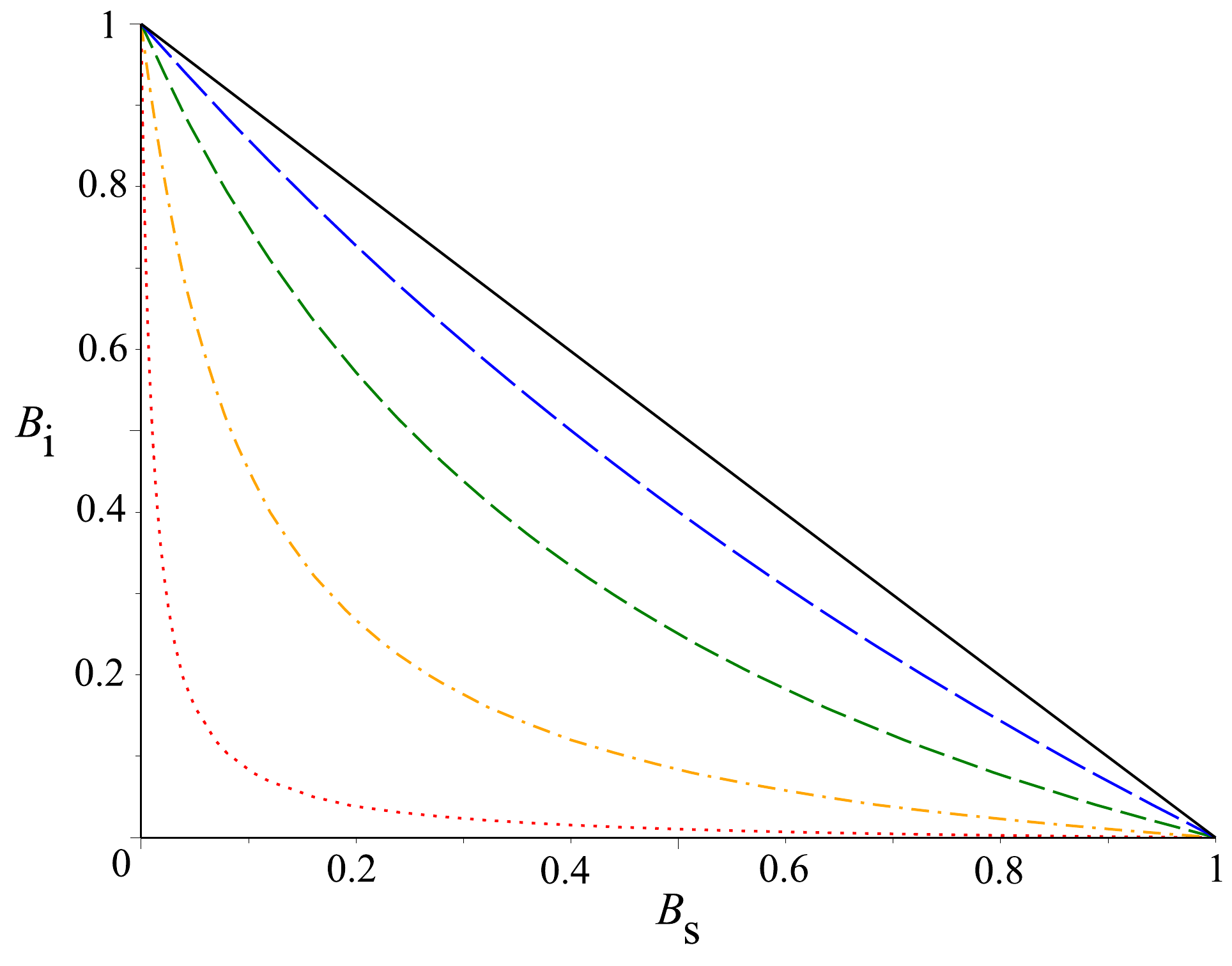}
 \caption{(Color online)
  Curves giving the boundaries between entangled and separable twin
  beams and determined according to Eq.~(\ref{18}) plotted
  in the plane spanned by the mean noise photon numbers $ B_{\rm s} $ and $ B_{\rm i} $
  assuming the mean photon-pair number $B_{\rm p}$ equal to
  $0.01$  [dotted (red) curve], $0.1$ [dash-dotted (yellow) curve], $0.5$ [dashed (green) curve], $2$ [long-dashed (blue) curve],
  and $B_{\rm p} = 100$ [solid black curve]. Entangled states are localized in the lower-left corner
  of the plane. The larger $ B_{\rm p} $, the larger the
  area containing entangled states.}
\label{fig3}
\end{figure}

According to Eq.~(\ref{negn}), all noiseless twin beams are
entangled. The more intense the noiseless twin beams are, the more
entangled the signal and idler fields are (see Fig.~\ref{fig1}).
The presence of noise in a twin beam can even completely destroy
entanglement, as the analysis of Eq.~(\ref{fneg1}) shows. Indeed,
the condition $ N > 0 $ for entanglement can be rewritten using
Eq.~(\ref{fneg1}) as follows:
\begin{equation} \label{neg_cond}       
B_{\rm p}[1-(B_{\rm s}+B_{\rm i})] > B_{\rm s}B_{\rm i} .
\end{equation}
Condition (\ref{neg_cond}) cannot be fulfilled for any value
of $ B_{\rm p} $ provided that $ B_{\rm s} + B_{\rm i} \ge 1 $.
Thus, the twin beam can be entangled only when
\begin{equation}  
 B_{\rm s} + B_{\rm i} < 1 \hspace{3mm} {\rm and} \hspace{3mm}
 B_{\rm p} > \frac{B_{\rm s}B_{\rm i}}{1-(B_{\rm s}+B_{\rm i})}  .
\label{18}
\end{equation}
The behavior of the negativity $ N $ of noisy twin beams dependent
 on the noise parameters $ B_{\rm s} $ and $ B_{\rm i} $
is illustrated in Fig.~\ref{fig2} for several values of the mean
photon-pair number $ B_{\rm p} $. It holds in general that the
greater the value of the mean photon-pair number $ B_{\rm p} $, the
greater the value of the negativity $ N $. This can be
explained as follows. The more intense twin beams, with their
thermal statistics, are effectively spread over a larger number of
the Fock states. This naturally results in the larger
effectively populated Hilbert spaces used to describe the
entanglement. The greater value of the negativity $ N $ means a
greater effective number of the paired modes building the
entanglement, i.e., a greater value of the entanglement
dimensionality, as defined in Sec.~V. Also, the greater the value
of the mean photon-pair number $ B_{\rm p} $, the larger the amount of
overall noise $ B_{\rm s} + B_{\rm i} $ acceptable in an entangled
twin beam (see Fig.~\ref{fig3}). The curves plotted in
Fig.~\ref{fig3} indicate that entanglement is more resistent to
noise when the noise is distributed in the signal and idler
fields asymmetrically. We note that separable states (i.e., with $
N=0 $) contain, in general, paired, signal, and idler noisy
contributions. However, the noisy contributions are sufficiently
strong to suppress the ``entangling power'' of the photon-pair
contribution and so the state effectively behaves as a classical
statistical mixture of the signal and idler fields.

\begin{figure}            
 \includegraphics[width=0.5\textwidth]{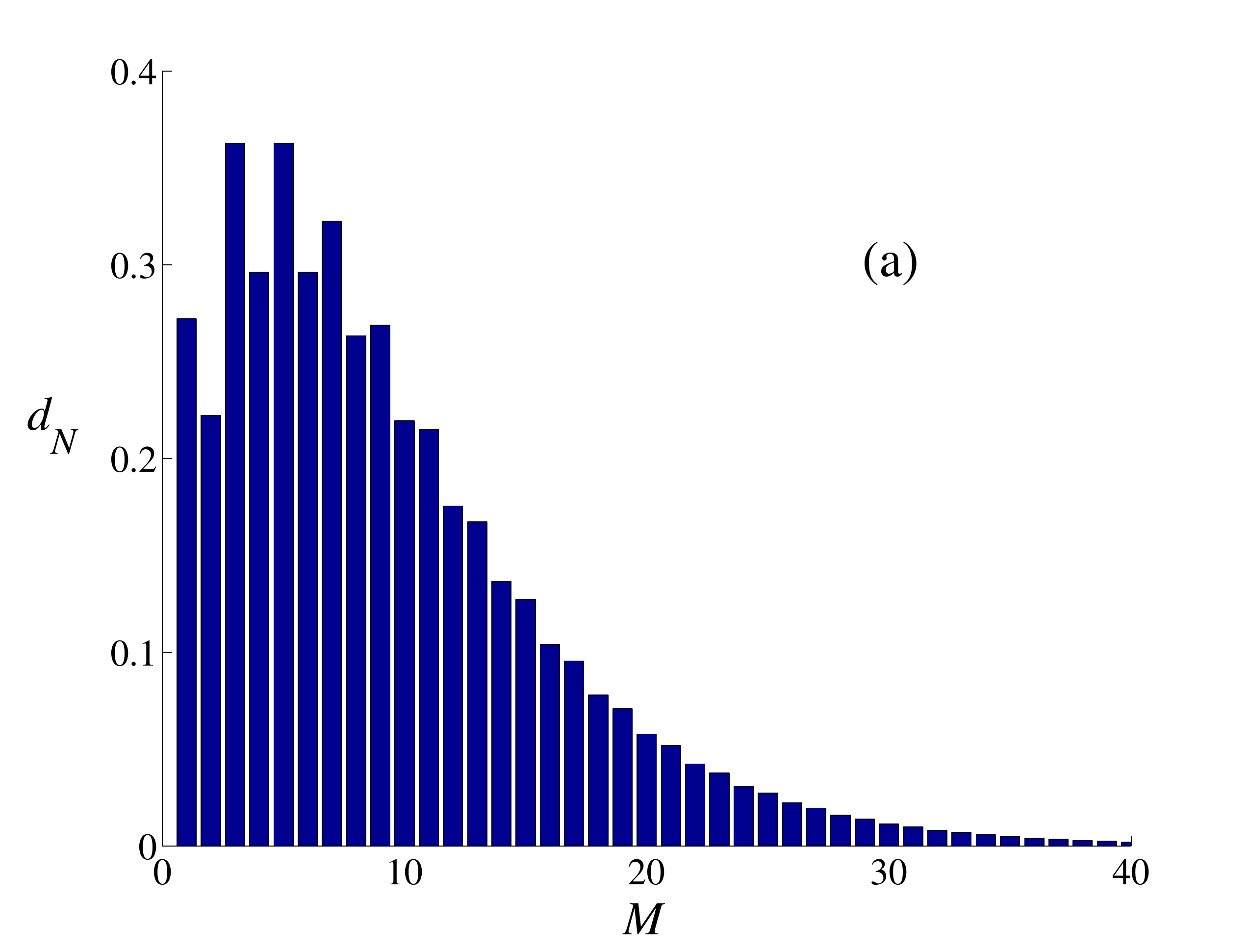}
 \includegraphics[width=0.5\textwidth]{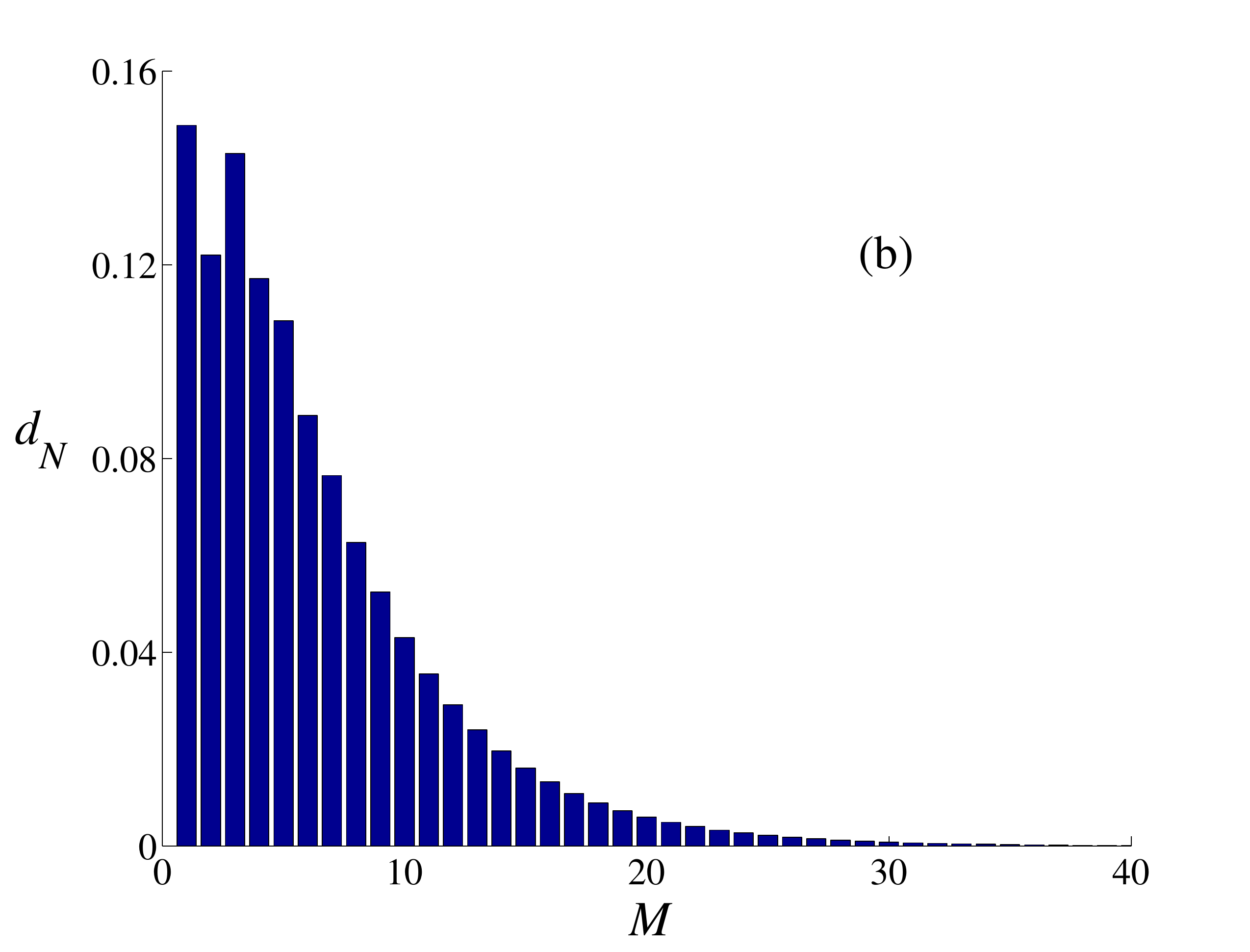}
 \caption{(Color online) Distribution $d_{N}$ of negativity $N$ given in Eq.~(\ref{distribution})
 assuming  $B_{\rm p} = 2 $ and (a) $ B_{\rm s}=B_{\rm i}=0$ and (b) $ B_{\rm s} = B_{\rm i} = 0.1 $.
 Note that $-d_N(M)$ corresponds to the sum of all the negative
 eigenvalues for the $(M+1$)--dimensional block of the partially
 transposed statistical operator $\hat\rho^\Gamma$. Thus, $d_{N}(M)$
 shows the internal structure of entanglement in the Liouville
 space.}
\label{fig4}
\end{figure}

The decomposition of the partially transposed statistical operator
$ \hat\rho^{\Gamma} $ into blocks in its matrix representation and
the fact that a block (subspace) with dimension $ M+1 $ describes
only states with up to $ M $ photons in the signal (and also
idler) field can be used to define the distribution $ d_N $ of the
negativity $ N $ fulfilling the normalization condition
\begin{eqnarray} 
  \sum_{M=1}^{\infty} d_N(M) = N.
\label{distribution}
\end{eqnarray}
For a given $M$, the element $ d_N(M) $ of this distribution is
given as the sum of the absolute values of the negative
eigenvalues belonging to the block of dimension $ M+1$. The
distribution $ d_N $ of the negativity provides insight into the
internal structure of entanglement. It tells us how entanglement
is distributed in the Liouville space of statistical operators.
Typical distributions $ d_N $ of the negativity for noiseless as
well as noisy twin beams are plotted in Fig.~\ref{fig4}.
A teeth-like structure occurs for smaller numbers $ M $ in noiseless
twin beams. Noise tends to suppress this structure, as is evident
from the comparison of the distributions $ d_N $ plotted in
Figs.~\ref{fig4}(a) and \ref{fig4}(b). We note that the densities of the
negativity have already been introduced for bipartite entangled
states composed of a qubit and continuum of
states~\cite{Luks2012,Perinova2014} as well as two continua of
states.

\section{Nonclassical depth of the twin beam}

\begin{figure}  
 \includegraphics[width=0.5\textwidth]{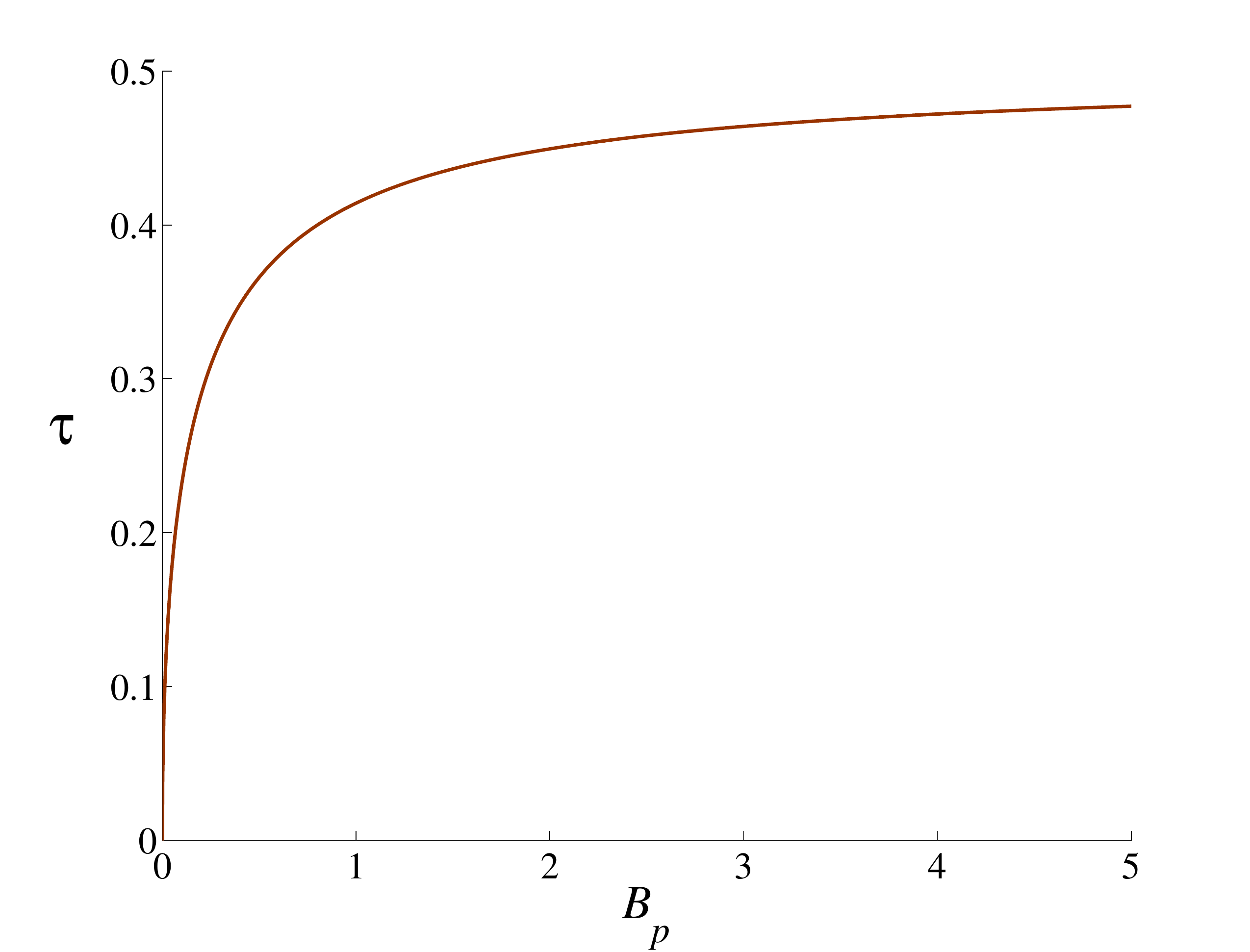}
 \caption{(Color online) Nonclassical depth $\tau$ given in Eq.~(\ref{nondepn}) as it
 depends on the mean photon-pair number $B_{\rm p}$ for noiseless twin beams, i.e.,
 $ B_{\rm s} = B_{\rm i} = 0 $.}
 \label{fig5}
\end{figure}

To quantify nonclassicality of the twin beam we apply the
nonclassical depth $\tau $~\cite{Lee91} derived from the threshold
value $s_{\rm th}$ of the ordering parameter at which the joint
signal-idler quasidistribution of integrated intensities becomes
nonnegative~\cite{Perina05,PerinaJr2013a}. We adopt the definition
$\tau = (1-s_{\rm th})/2 $. We note that the joint signal-idler
quasidistribution of integrated intensities attains negative
values for $ 1 \ge s > s_{\rm th}$ for which $ \tau > 0 $. The
threshold value $ s_{\rm th} $ can easily be obtained from the
condition $ \langle[\Delta (W_{\rm s}-W_{\rm i})]^2\rangle =0 $,
which determines the point of the transition between quantum and
classical single-mode twin beams~\cite{PerinaJr2013a}. This
results in the following formula for the nonclassical depth $ \tau
$:
\begin{equation}\label{nondep}  
 \tau = \frac{1}{2}\left[\sqrt{(B_{\rm s}-B_{\rm i})^2+4B_{\rm p}(B_{\rm p}+1)} - 2B_{\rm p}-B_{\rm s}-B_{\rm i})\right].
\end{equation}
Assuming noiseless twin beams, Eq.~(\ref{nondep}) simplifies to
\begin{equation}\label{nondepn} 
\tau = \sqrt{B_{\rm p}(B_{\rm p}+1)}-B_{\rm p}.
\end{equation}
According to Eq.~(\ref{nondepn}), all noiseless twin beams are
nonclassical. The greater the mean photon-pair number $ B_{\rm p}
$, the greater the value of the nonclassical depth $ \tau $ (see
Fig.~\ref{fig5}). This depth $ \tau $ reaches its greatest value,
1/2, in the limit of an infinitely intense twin beam ($ B_{\rm p}
\rightarrow \infty $). We note that $ \tau = 1/2 $ corresponds to
symmetrical ordering of the field operators.

\begin{figure}  
 \includegraphics[width=0.5\textwidth]{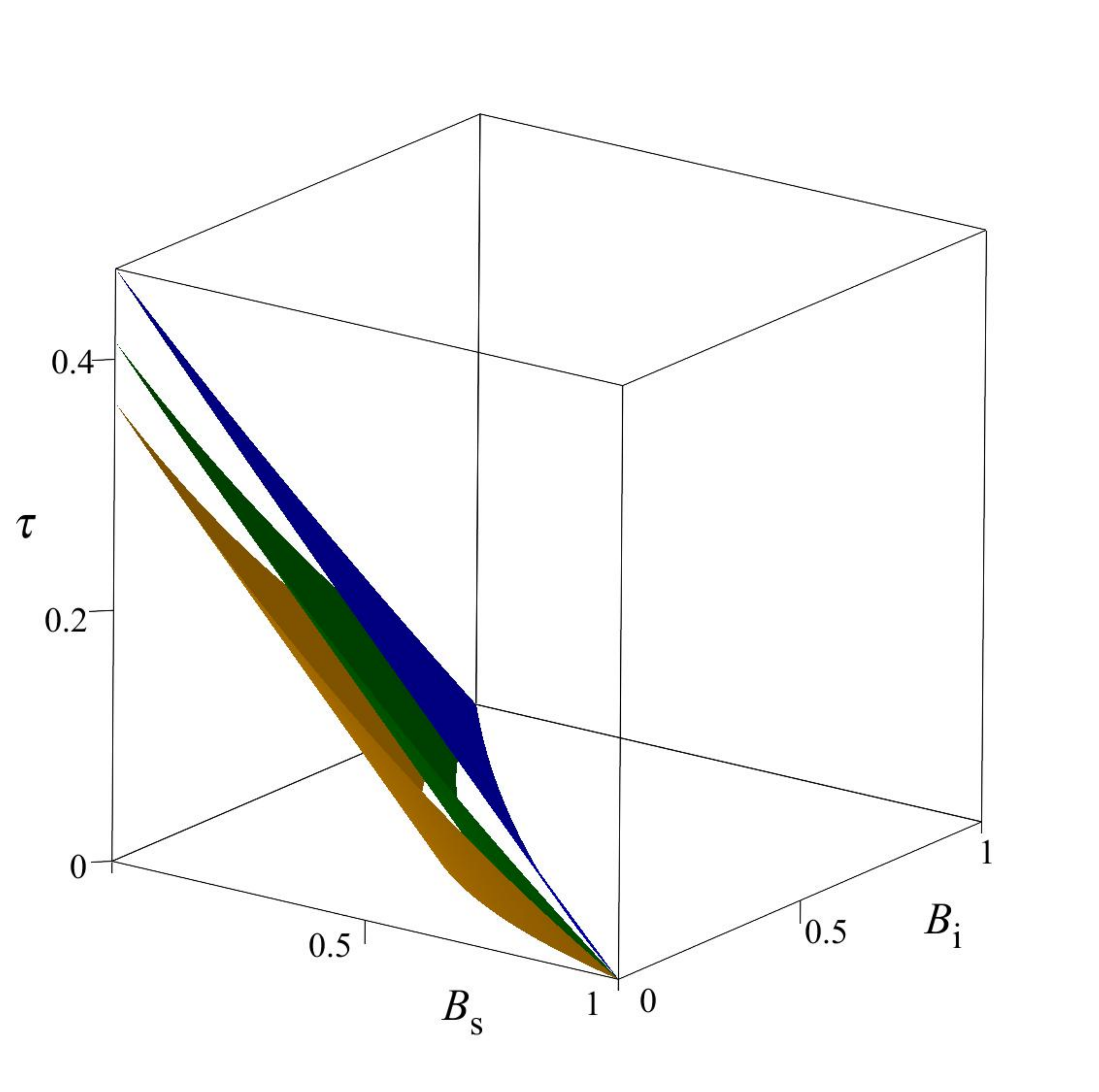}
 \caption{(Color online) Nonclassical depth $\tau$ given in Eq.~(\ref{nondep}) as a
  function of the mean noise photon numbers $B_{\rm s}$ and $B_{\rm i}$ for the mean
  photon-pair number $B_{\rm p}$ equal to $0.1$ [bottom, light-gray (yellow) area],
  $0.5 $ [gray (green) area], $4 $ [top, dark-gray (blue) area]. The greater the value of $ B_{\rm p} $,
  the greater the value of~$ \tau $.}
\label{fig6}
\end{figure}

On the other hand, and according to Eq.~(\ref{nondep}), noise only
degrades nonclassical behavior of a twin beam, as documented in
Fig.~\ref{fig6}. If the noise is equally distributed in the signal
and idler fields ($ B_{\rm s} = B_{\rm i} $), the nonclassical
depth $\tau $ determined along Eq.~(\ref{nondep}) gives the mean
number $ B_{\rm s}+B_{\rm i} $ of noise photons needed for
suppressing the nonclassicality of the twin beam. So, the larger the
value of the nonclassical depth $\tau $ is, the more nonclassical
the field is. On the other hand, formal application of
Eq.~(\ref{nondep}) to classical noisy twin beams results in
negative values of the nonclassical depth $\tau $. Their absolute
value $ |\tau| $ can be considered a measure of classicality of
noisy twin beams in the sense that it quantifies the mean number
of photon pairs needed to transform a classical twin beam into
the classical-quantum boundary $ \tau =0 $.

Condition $ \tau = 0 $ for the transition from quantum to
classical twin beams applied to Eq.~(\ref{nondep}) results in the
same relation among parameters $ B_{\rm p} $, $ B_{\rm s} $, and $
B_{\rm i} $ as derived in Eq.~(\ref{neg_cond}) for the
boundary between entangled and separable twin beams. Thus,
entangled twin beams are nonclassical, whereas separable twin beams
are classical. This means that nonclassical twin beams may contain
on average only less than one noise photon ($ B_{\rm s} + B_{\rm
i} < 1 $). We note that inequality (\ref{neg_cond}) represents the
Simon criterion for nonclassicality of Gaussian states as shown in Ref.~\cite{Perina11}.

\begin{figure}  
 \includegraphics[width=0.5\textwidth]{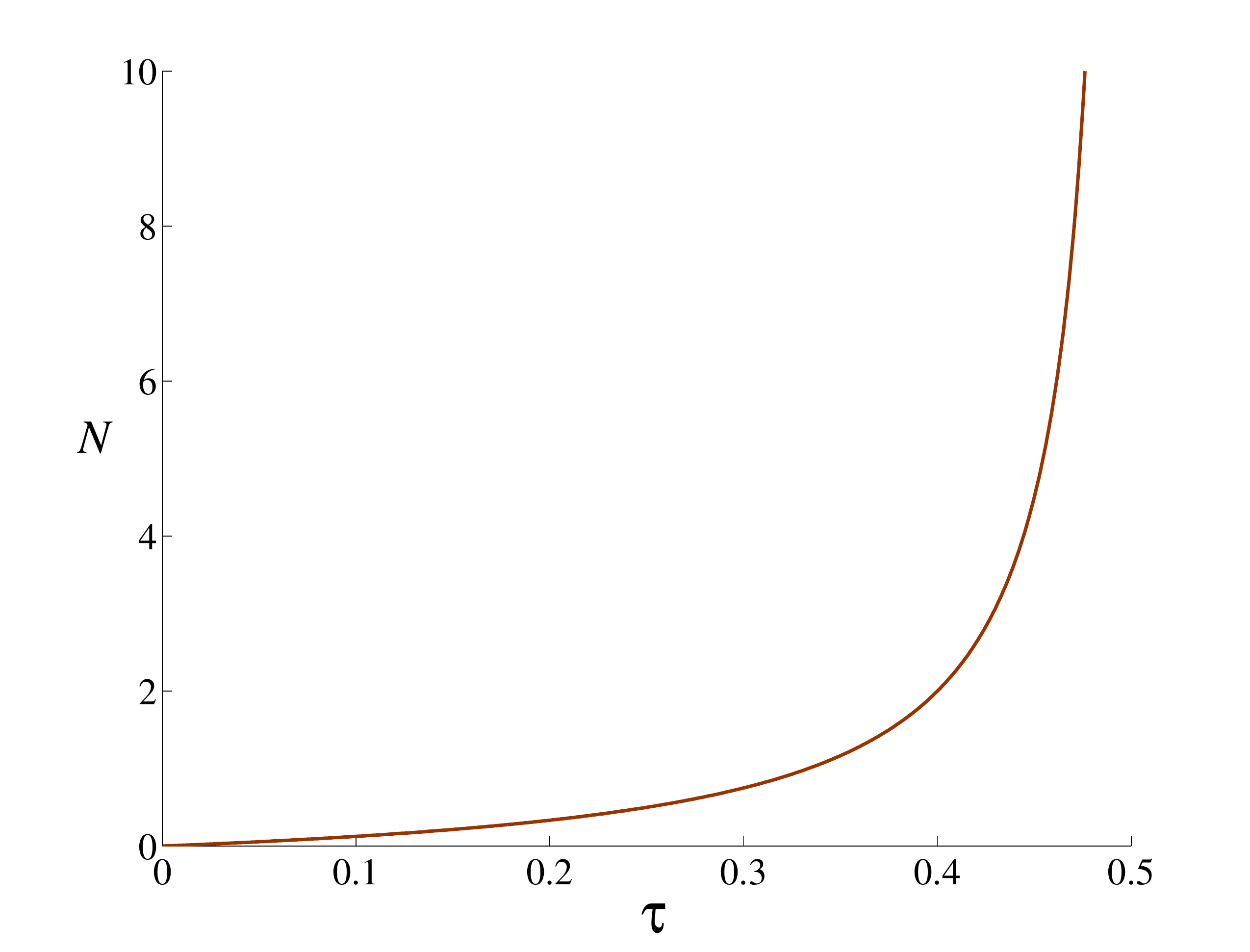}
 \caption{(Color online) Negativity $ N $ as a function of the nonclassical depth $ \tau $, according to Eq.~(\ref{22}).}
 \label{fig7}
\end{figure}

Comparison of Eqs.~(\ref{negn}) and (\ref{nondepn}) made for
noiseless twin beams reveals a simple relation between the
negativity $ N $ and the nonclassical depth $ \tau $:
\begin{equation}    
 N = \frac{\tau}{1-2\tau}.
\label{22}
\end{equation}
Direct calculation based on Eqs.~(\ref{fneg1}) and (\ref{nondep})
then confirms that relation (\ref{22}) holds even for a general
noisy twin beam. We thus have a one-to-one correspondence between
the value of the negativity $ N $ and the value of the nonclassical depth $ \tau $.
Moreover, according to Eq.~(\ref{22}) the negativity $ N $ is an
increasing function of the nonclassical depth $ \tau $, and vice
versa (see Fig.~\ref{fig7}). There exists a deep physical reason
for this correspondence. The nonlinear process emits photons in
pairs into the signal and idler fields, which creates entanglement
between these fields. It is this entanglement that gives rise to
nonclassical properties of twin beams, as the classical statistical
optics is unable to describe pairing of photons appropriately.

\section{Dimensionality of the twin beam}

Three different numbers are needed to determine the dimensionality of
a general noisy twin beam. The dimensionality $ K_{\rm ent} $ of
entanglement gives the number of degrees of freedom constituting
the entangled (paired) part of the twin beam. We also need
additional degrees of freedom to characterize the noisy parts of
the twin beam. As the amount of noise is, in general, different in
the signal and idler fields, we have independent participation
ratios $ R_{\rm s} $ and $ R_{\rm i} $ for both fields. The
entanglement dimensionality $ K_{\rm ent} $ for bipartite states
with axisymmetric statistical operators can be given in terms of
the negativity $ N $ by a simple formula~\cite{Eltschka13}:
\begin{equation}\label{Ndim}        
 K_{\rm ent}(\hat\rho) = 2N(\hat\rho) + 1=\vert\vert\hat\rho^{{\Gamma}}\vert\vert_{1}.
\end{equation}
Strictly speaking, it is the least integer $\ge K_{\rm ent}$ that
gives a lower bound to the number of entangled dimensions between
entangled subsystems (paired modes) of
$\hat\rho$~\cite{Eltschka13}. According to Eq.~(\ref{Ndim}),
the entanglement dimensionality $ K_{\rm ent} $ equals 1 for separable
states ($ N=0 $). It linearly increases with the negativity $ N $.
As the noise described by the mean noise photon numbers $ B_{\rm
s} $ and $ B_{\rm i} $ decreases the values of the negativity $ N
$, it also decreases the values of the entanglement dimensionality $
K_{\rm ent} $. We note that, for pure states, the Schmidt number
is also a good quantifier of the entanglement dimension $ K_{\rm ent}
$~\cite{Law04,Gatti2012,Chekhova2015}. The Schmidt decomposition
of pure states accompanied by convex optimization can even be
applied for quantifying the entanglement dimension of mixed
entangled states~\cite{Horodecki09review}.

On the other hand, the noise present in the signal and idler
fields requires additional degrees of freedom for its description.
These degrees of freedom are, together with those reserved for
describing entanglement, determined by the participation ratios $
R_{\rm s} $ and $ R_{\rm i} $ derived from the signal- and
idler-field reduced statistical operators $ \hat\rho_{\rm s} $ and
$ \hat\rho_{\rm i} $, respectively~\cite{Gatti2012,Horoshko2012}:
\begin{equation}                
 R_{a} = \frac{1}{\mathrm{Tr}_{a}[\hat\rho_{a}^{2}]},
  \hspace{5mm} a={\rm s,i}.
\label{24}
\end{equation}
Equation~(\ref{rho1}), giving the matrix elements of the statistical
operator $ \hat\rho $, guarantees a diagonal form of the reduced
statistical operators $ \hat\rho_{\rm s} $ and $ \hat\rho_{\rm i}
$ of the signal and idler fields, respectively. In this case,
Eq.~(\ref{24}) can be rewritten in the form
\begin{equation}\label{Kfin}    
 R_{\rm s} = \frac{1}{ \sum_j \rho_{{\rm s},jj}^2}.
\end{equation}
Using Eq.~(\ref{rho1}) the matrix elements $\rho_{{\rm s},jj} $
can be written as
\begin{equation}\label{rhoii}   
 \rho_{{\rm s},jj} =\frac{1}{\tilde B_{\rm s}} \left[\left(1-\frac{\tilde B_{\rm i}}{\tilde K}\right)
  +\frac{\vert D_{12}\vert^2}{\tilde K \tilde B_{\rm s}}\right]^j.
\end{equation}
Substituting Eq.~(\ref{rhoii}) into Eq.~(\ref{Kfin}) we obtain a
simple formula for the participation ratio $ R_{\rm s} $:
\begin{equation}\label{K}       
 R_{\rm s} = 2(B_{\rm p} + B_{\rm s}) + 1.
\end{equation}
The same considerations made for the signal field apply also to
the idler field.

To find the relation between the entanglement dimensionality $
K_{\rm ent} $ and the participation ratios $ R_{\rm s} $ and $
R_{\rm i} $ we consider for a while the noiseless twin beams in
pure states. In this case, the elements $ \hat\rho_{{\rm s},jj} $
of the reduced statistical operator $ \hat\rho_{\rm s} $, written
in Eq.~(\ref{rhoii}), immediately give the squared Schmidt
coefficients~\cite{Pires09}. Combining Eqs.~(\ref{negn}),
(\ref{Ndim}), and (\ref{K}) we arrive at the formula
\begin{equation}            
 K_{\rm ent} = R_{\rm s} + \sqrt{R_{\rm s}^2-1} .
\label{28}
\end{equation}
Equation~(\ref{28}) shows that, excluding weak noiseless twin
beams, $ K_{\rm ent} \approx 2R_{\rm s} $. This means that the
definitions of the entanglement dimensionality and participation ratio
set different boundaries for the Schmidt coefficients $c_{j}$
included in the approximative description of a noiseless twin beam
with the wave function
\begin{equation}    
  |\psi\rangle =\sum_{j=0}^{j_{\rm max}}  c_j |j\rangle_{\rm s} |j\rangle_{\rm
  i}. \label{psi}
\end{equation}
Using Eq.~(\ref{rho1}), the coefficients $ c_j $ in
Eq.~(\ref{psi}) are obtained in the form
\begin{equation}            
 c_j = \sqrt{\frac{{B_{\rm p}}^{j}}{(B_{\rm p}+1)^{j+1}}},
\label{30}
\end{equation}
which is in agreement with the thermal photon-number statistics of
the signal (or idler) field. We note that the ratio $ c_{K_{\rm
ent}-1} / c_{R_{\rm s} -1} $ of boundary coefficients is given by
the expression $ [B_{\rm p}/(1+B_{\rm p})]^{B_{\rm p}+1} $. When
$B_{\rm p} \rightarrow \infty $ $c_{K_{\rm ent}-1}/c_{R_{\rm s}-1}
\rightarrow 1/e$.

To compare the values of entanglement dimensionality and the
participation ratio for general twin beams we have to eliminate
the effect of different boundaries set by different definitions,
as revealed by considering the pure states. Using the formulas
derived for noiseless twin beams, we introduce the modified
entanglement dimensionality $ \tilde K_{\rm ent} $ as follows:
\begin{equation}   
 \tilde K_{\rm ent} = \frac{2B_{\rm p}+1}{2B_{\rm p}+1 + 2\sqrt{B_{\rm p}^2+B_{\rm p}}}\,
  K_{\rm ent} .
\label{31}
\end{equation}
Definition (\ref{31}) of the modified entanglement dimensionality
$ \tilde K_{\rm ent} $ guarantees that the values of modified
entanglement dimensionality $ \tilde K_{\rm ent} $ and
participation ratios $ R_{\rm s} $ and $ R_{\rm i} $ of noiseless
twin beams are equal.

\begin{figure} [t!]
 \includegraphics[width=0.5\textwidth]{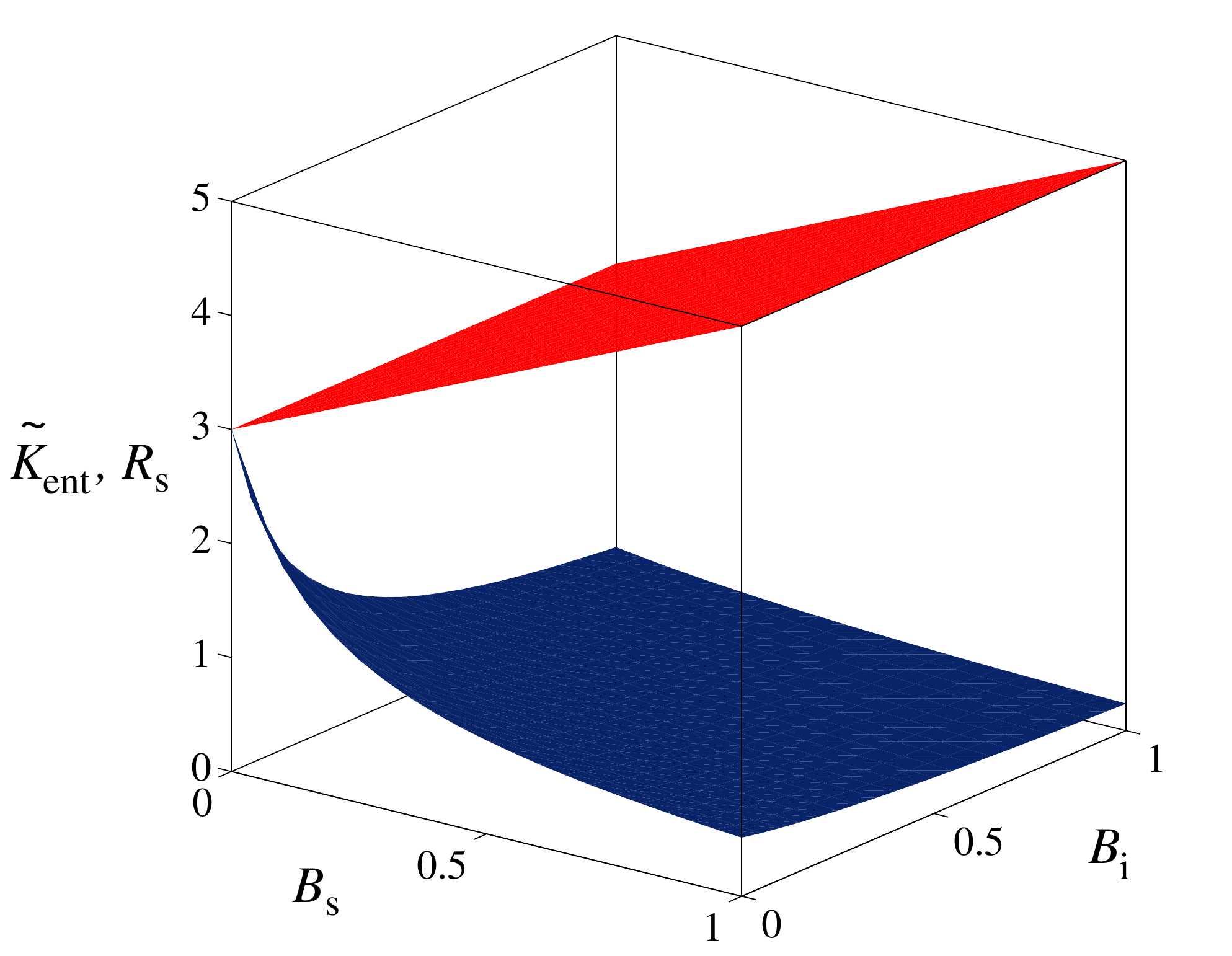}
 \caption{(Color online) Modified entanglement dimensionality $ \tilde K_{\rm ent}$
 given in Eq.~(\ref{31}) [lower, dark-gray (blue) area] and
 signal-field participation ratio $ R_{\rm s} $ given in Eq.~(\ref{K}) [upper, gray (red) area]
 as they depend on the mean noise photon numbers
 $B_{\rm s}$ and $B_{\rm i}$ assuming the mean photon-pair number $B_{\rm p}= 1$.}
\label{fig8}
\end{figure}

The values of the modified dimensionality $ \tilde K_{\rm ent} $ of
entanglement and the signal-field participation ratio $ R_{\rm s} $
are compared in Fig.~\ref{fig8} for the mean photon-pair number $
B_{\rm p} =1 $. Whereas the values of the modified entanglement
dimensionality $ \tilde K_{\rm ent} $ decrease with increasing
values of the mean noise photon numbers $ B_{\rm s} $ and $ B_{\rm i}
$, the values of the signal-field participation ratio $ R_{\rm s} $
increase with increasing values of the mean signal-field noise
photon number $ B_{\rm s} $. We note that the values of the
signal-field participation ratio $ R_{\rm s} $ are greater than
those of the modified entanglement dimensionality $ \tilde K_{\rm
ent} $ even for $ B_{\rm s} = 0 $, as the presence of noise in the
idler field ($ B_{\rm i}> 0 $) degrades entanglement.

\begin{figure}  
 \includegraphics[width=0.5\textwidth]{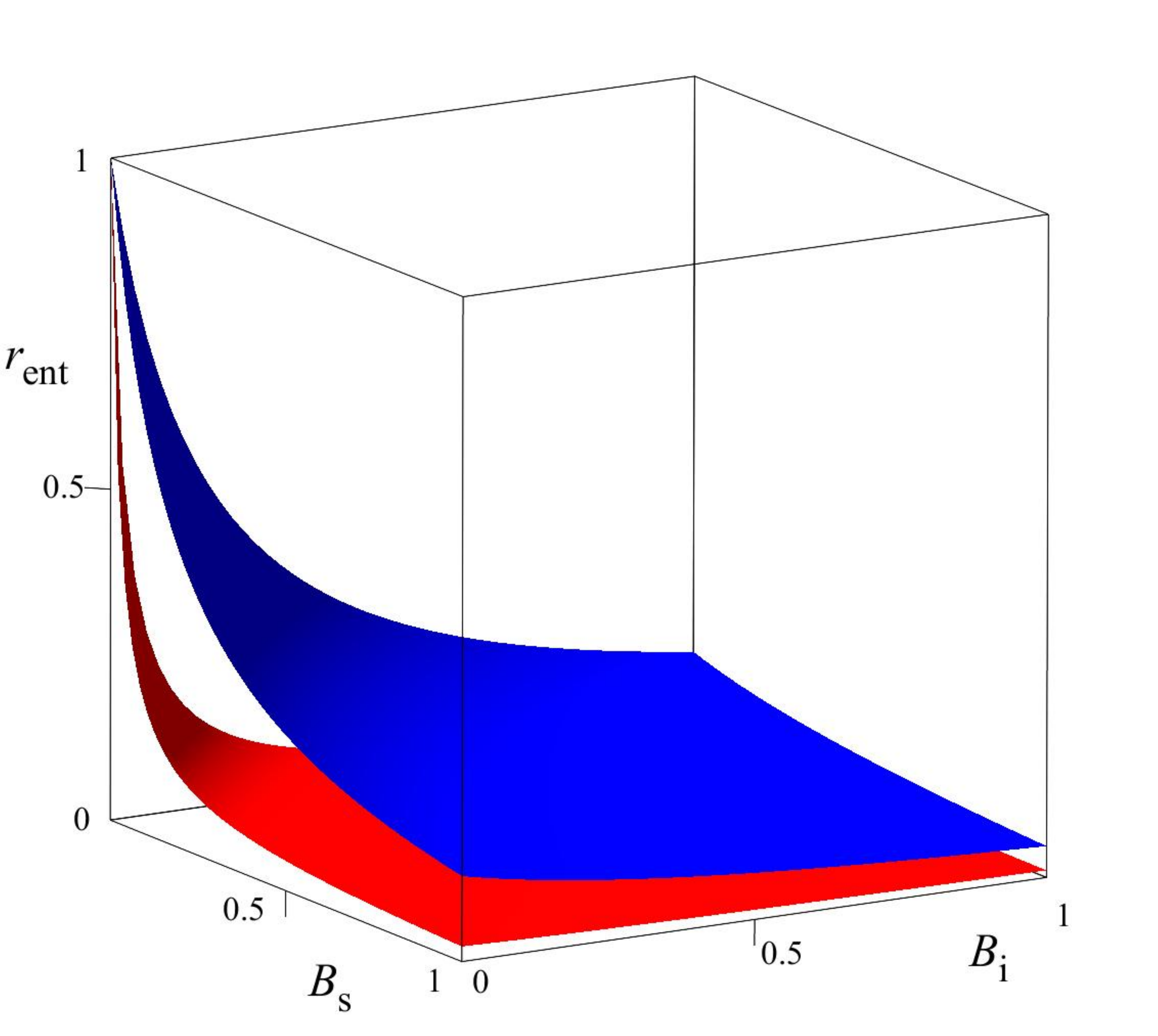}
 \caption{(Color online) Coefficient $ r_{\rm ent} $ given in
  Eq.~(\ref{32}) versus the mean noise photon numbers $B_{\rm s}$ and $B_{\rm i}$ for the mean
  photon-pair number $B_{\rm p}$ equal to $1$ [upper, dark-gray (blue) area],
  and $ 10 $ [lower, gray (red) area].}
\label{fig9}
\end{figure}

The relative contribution of the degrees of freedom used for describing
entanglement in a twin beam is an important characteristic. This
contribution can be quantified via the coefficient $ r_{\rm ent} $
defined as follows:
\begin{equation}  
 r_{\rm ent} = \frac{2\tilde K_{\rm ent}}{R_{\rm s} + R_{\rm i}}.
\label{32}
\end{equation}
As shown in Fig.~\ref{fig9}, the greater the values of the mean noise
photon numbers $ B_{\rm s} $ and $ B_{\rm i} $, the smaller the
values of the coefficient $ r_{\rm ent} $. The comparison of surfaces
of the coefficient $ r_{\rm ent} $ drawn for the mean photon-pair
numbers $ B_{\rm p}=1 $ and $ B_{\rm p}=10 $ in Fig.~\ref{fig9}
reveals seemingly paradoxical behavior. The values of the coefficient
$ r_{\rm ent} $ decrease with increasing values of the mean
photon-pair number $B_{\rm p}$. This behavior, however, naturally
originates in fragility of entanglement with respect to the noise.
More intense twin beams (with greater values of $ B_{\rm p} $) are
less resistant to a given amount of noise compared to
low-intensity twin beams. This is explained by the larger
dimensions of the effectively populated Hilbert spaces of more
intense twin beams and, thus, the more complex structures of their
entanglement. As a consequence, relatively higher numbers of
degrees of freedom serving to describe entanglement in more
intense noiseless twin beams are ``released'' by the noise and
enlarge the noise parts of twin beams.

\begin{figure} [b!]         
 \includegraphics[width=0.5\textwidth,height=6cm]{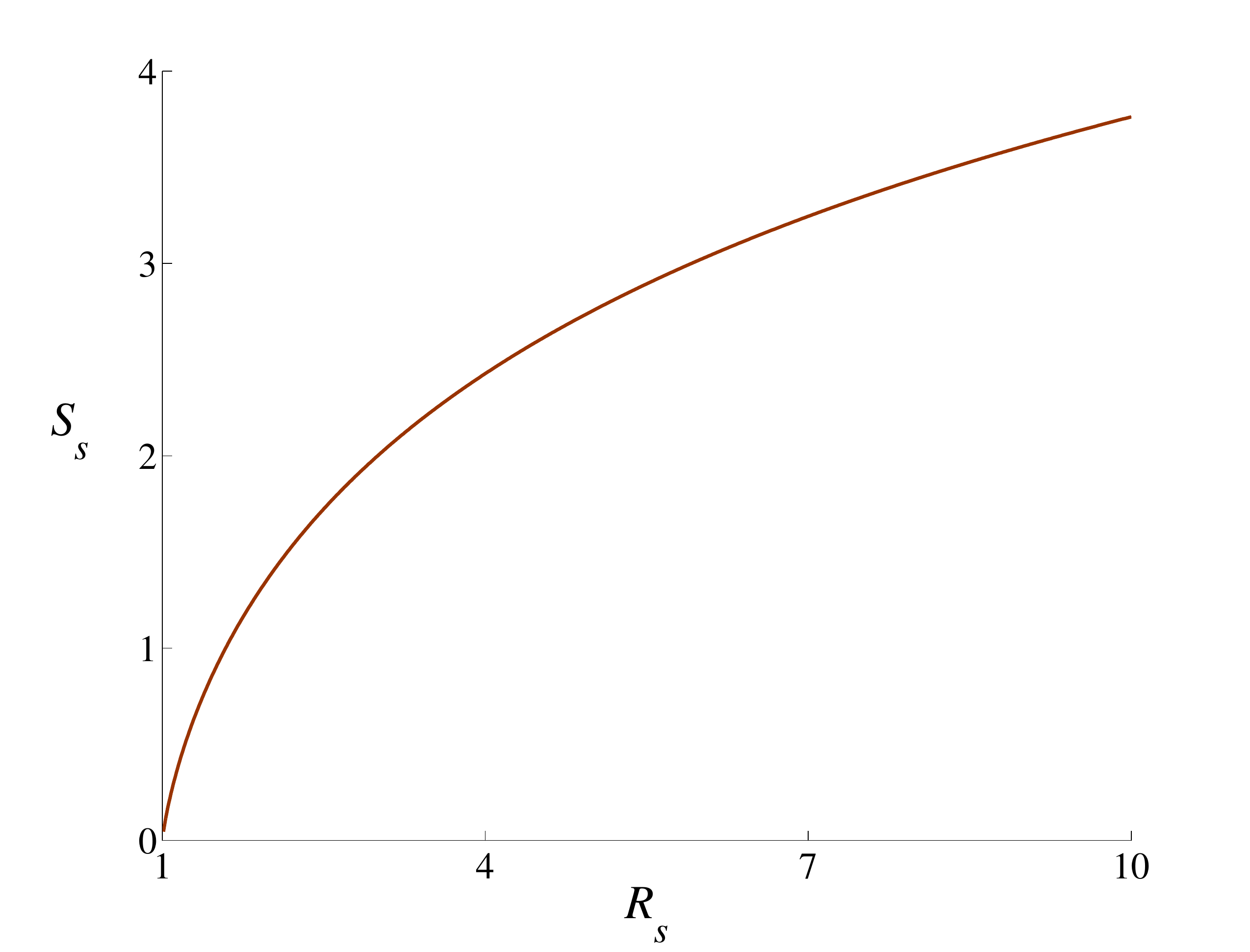}
 \caption{(Color online) von Neumann entropy $S_{\rm s}$ as a function of the participation ratio
 $R_{\rm s}$ according to Eq.~(\ref{funSofK}).}
\label{fig10}
\end{figure}

Alternatively to the participation ratio $ R $, we may apply the
von Neumann entropy $S$ of a reduced statistical operator. Taking
into account the diagonal form of the signal-field reduced
statistical operator $ \hat\rho_{\rm s} $ with the elements
written in Eq.~(\ref{rhoii}), the signal-field entropy $ S_{\rm s}
$ is in general determined along the formula
\begin{equation}\label{S}       
 S_{\rm s} = - \mathrm{Tr}(\hat \rho_{\rm s} \ln \hat \rho_{\rm s})=  -\sum_j \rho_{{\rm s},jj}\ln(\rho_{{\rm s},jj}).
\end{equation}
Considering the specific form of matrix elements $ \rho_{{\rm
s},jj} $ given in Eq.~(\ref{rhoii}), the formula for entropy $
S_{\rm s} $ attains the form
\begin{eqnarray}\label{Sfin}       
 S_{\rm s} &=& (1+B_{\rm p}+B_{\rm s}) \ln(1+B_{\rm p}+B_{\rm s}) \nonumber \\
 & & \mbox{} -(B_{\rm p}+B_{\rm s})\ln(B_{\rm p}+B_{\rm s});
\end{eqnarray}
$ \ln $ stands for natural logarithm. Combining Eqs.~(\ref{Kfin})
and (\ref{Sfin}), the entropy $ S_{\rm s} $ is revealed as an
increasing function of the participation ratio $ R_{\rm s} $:
\begin{equation}\label{funSofK}    
 S_{\rm s} = \frac12\left[(R_{\rm s}+1)\ln (R_{\rm s}+1) -(R_{\rm s}-1)\ln(R_{\rm s}-1)\right] - 1.
\end{equation}
Analogous formulas for the idler-field entropy $ S_{\rm i} $ can
easily be derived. The general dependence of entropy $ S_{\rm s} $
on the participation ratio $ R_{\rm s} $ is plotted in
Fig.~\ref{fig10}.
We would like to note that the entropy $S$ serves as a good measure of the
entanglement for pure states.

\section{Twin beam composed of $ M $ modes}

In real experiments, twin beams are rarely composed of only one
paired spatiotemporal mode~\cite{PerinaJr2013a,Allevi2013}. We
note that a twin beam composed of one paired mode represents an
ideal field from the experimental point of view~\cite{Perez14}.
For this reason, we consider a multimode twin beam containing $ M
$ independent identical single-mode twin beams. Its statistical
operator $ \hat\rho_M $ is given as $\hat\rho_M = \otimes_M
\hat\rho $ using the statistical operator $ \hat\rho $ written in
Eq.~(\ref{rhofor}). There are four parameters characterizing the
twin beam: number $ M $ of modes, mean photon-pair number $ B_{\rm
p} $, mean signal-field noise photon number $ B_{\rm s} $, and
mean idler-field noise photon number $ B_{\rm i} $. We note that
such an $ M $-mode twin beam represents a good approximation of a
real twin beam when all spatiotemporal modes participating in the
nonlinear interaction are detected.

The considered physical quantities behave differently with respect
to the number $ M $ of modes. It has been shown in
Refs.~\cite{Perina05} and \cite{PerinaJr2013a} that the nonclassical depth $
\tau $ does not depend on the number $ M $ of modes. On the other
hand, the multimode negativity $ N_M $, $ N_M = (1+2N)^M $, as
well as the participation ratios $ R_{M,a} $, $ R_{M,a} = R_a^M $
for $ a=s,i $, are multiplicative. We note that the form of the
multimode negativity originates in the multiplicative property of the
trace norm and its relation to the negativity expressed in
Eq.~(\ref{neg})~\cite{Vidal02}. In fact, the multimode negativity
$ N_M $ coincides with the entanglement dimensionality $ K_{\rm
ent} $ defined in Eq.~(\ref{Ndim}) for a single-mode twin beam.
The multimode entropies $S_{M,a}$, $a={\rm s,i}$, are then
additive. To reveal similar relations among the studied quantities
as has been done for single-mode twin beams, we have to define
suitable quantities derived from those considered above. Defining
the logarithmic negativity $ N_M^{\rm log} \equiv \ln(N_M) $ and
the logarithmic participation ratios $ R_{M,a}^{\rm log} \equiv
\ln(R_{M,a})$, $a={\rm s,i}$, we replace the multiplicative
quantities with the additive ones. Introducing the logarithmic
negativity $ {\mathcal N} $, logarithmic participation ratios $
{\mathcal R}_{a}^{\rm log} $, and entropies $ {\mathcal S}_a $
related per one mode,
\begin{eqnarray}  
 {\mathcal N} &=& \frac{N_M^{\rm log}}{M} = \ln (1+2N) , \nonumber \\
 {\mathcal R}_a &=& \frac{R_{M,a}^{\rm log}}{M} = \ln (R_a) , \nonumber \\
 {\mathcal S}_a &=& \frac{S_{M,a}}{M} = S_a,
\label{36}
\end{eqnarray}
with $a={\rm s,i}$, we reveal the suitable quantities. The
quantities defined in Eq.~(\ref{36}) together with the
nonclassical depth $ \tau $ behave qualitatively in the same way
as those defined for single-mode twin beams discussed above.
Especially, the logarithmic negativity $ {\mathcal N} $ per mode
is an increasing function of the nonclassical depth $ \tau $. Also,
the entropy $ {\mathcal S}_a $ per mode is an increasing function of
the logarithmic participation ratio $ {\cal R}_a $ per mode,
$a={\rm s,i}$.

\section{Experimental multimode twin beams}

Real experimental multimode twin beams have a more complex
structure than that discussed in
Sec.~VI~\cite{Haderka2005a,PerinaJr2013a,Allevi2013}. The reason
is that the spatiotemporal modes of twin beams are shared by the
signal and idler fields and so they can be broken before or during
the detection owing to spectral and/or spatial filtering. As a
consequence, real multimode twin beams are composed of three
components~\cite{PerinaJr2013a,PerinaJr2012a}. A paired component
describes photons embedded in spatiospectral modes detected by
both signal- and idler-field detectors. A noise signal (idler)
component then describes photons occurring in signal (idler)
spatiotemporal modes that originate in filtering of the idler
(signal) field. If we assume for simplicity that the paired
component is ideal, i.e., without noise, we need six parameters to
describe a real twin beam. Each component is characterized by the
number $ M $ of modes and mean photon-pair (or noise photon)
number $ B $. The statistical operator $ \hat\rho_E $ of the
experimental twin beam can be expressed as
\begin{equation}  
 \hat\rho_E= \bigotimes_{M_{\rm p}} \hat\rho_{\rm p}
  \bigotimes_{M_{\rm s}} \hat\rho_{n,s} \bigotimes_{M_{\rm i}}
  \hat\rho_{n,{\rm i}}
\label{37}
\end{equation}
using single-mode statistical operators $ \hat\rho_{\rm p} $, $
\hat\rho_{n,s} $, and $ \hat\rho_{n,i} $ of the photon-pair, noise
signal, and noise idler components. In Eq.~(\ref{37}), $M_{\rm
p}$, $M_{\rm s}$, and $M_{\rm i}$ give the numbers of equally
populated modes with the mean numbers $B_{\rm p}$, $B_{\rm s}$,
and $B_{\rm i}$ of photon pairs per mode, respectively.

Entanglement in the experimental twin beam is created only by its
paired component and as such it can be quantified by the
logarithmic negativity $ N_{M_{\rm p}}^{\rm log} $ introduced in
Sec.~VI. The noise components do not contribute to entanglement on
one side, and they do not degrade entanglement on the other side. This
is qualitatively different from the case of multimode twin beams
discussed in Sec.~VI and containing noise in paired
spatiotemporal modes.

Nonclassicality can be quantified by a multimode generalization of
nonclassical depth $ \tau_E $ introduced in Ref.~\cite{Lee91} for
a single-mode field. In a multimode twin beam, we may first
determine the standard nonclassical depths $ \tau_n $ for each
single-mode field, included either in the paired part of the twin
beam or in the noisy signal and idler parts of the twin beam. Then
we can take either $ \max_n(\tau_n) $ or $ \sum_n \tau_n $ to
quantify the multimode nonclassical depth $ \tau_E $. In the first
case, the nonclassical depth $ \tau_E $ of the experimental
multimode twin beam is just given by the nonclassical depth $ \tau
$ of a paired mode. The second case is physically more interesting,
as the value of $ \tau_E $ is linearly proportional to the minimum
amount of additional noise needed to conceal nonclassicality of
the multimode state. In this case, we have, for the experimental
multimode twin beams,
\begin{equation}  
 \tau_E = M_{\rm p} \tau .
\label{38}
\end{equation}
Using the logarithmic negativity $ N^{\rm log}_{M_p} $ defined in
Sec.~VI and the nonclassical depth $ \tau_E $, one-to-one
correspondence between the entanglement and the nonclassicality is
obtained also for $ M $-mode twin beams.

On the other hand, the concept of weak
nonclassicality~\cite{Arvind1997,Arvind1998,DodonovBook} is also
useful for the experimental multimode twin beams considered to be
composed of one effective paired (macro)mode. The joint
quasidistribution $ P_W $ of the integrated intensities $ W_s $ and $
W_i $ of the signal and idler fields, respectively, describes the
properties of this effective paired mode~\cite{Perina1991Book}. As
no information about the phase is encoded in this simplified
effective description, we may only determine the nonclassical
intensity depth $ \tau_W $ quantifying nonclassicality, which
demonstrates itself by negative values of the marginal
quasidistribution of integrated intensities. We have to emphasize
that the nonclassical intensity depth $ \tau_W $ is only a
nonclassicality witness or parameter, which reveals
nonclassicality solely in photon-number statistics. Contrary to
this, the nonclassical depth $ \tau $ is a genuine and commonly
used nonclassicality measure. We note that te standard
nonclassicality quantified by $\tau$ reveals both strongly and
weakly nonclassical states~\cite{Arvind1997,Arvind1998}. From this
point of view $\tau$ is a \emph{strong} tool or criterion. On the
other hand, $\tau_W$ detects only strongly nonclassical states;
i.e., it is a \emph{weak} tool.

\begin{figure} 
 \includegraphics[width=0.5\textwidth]{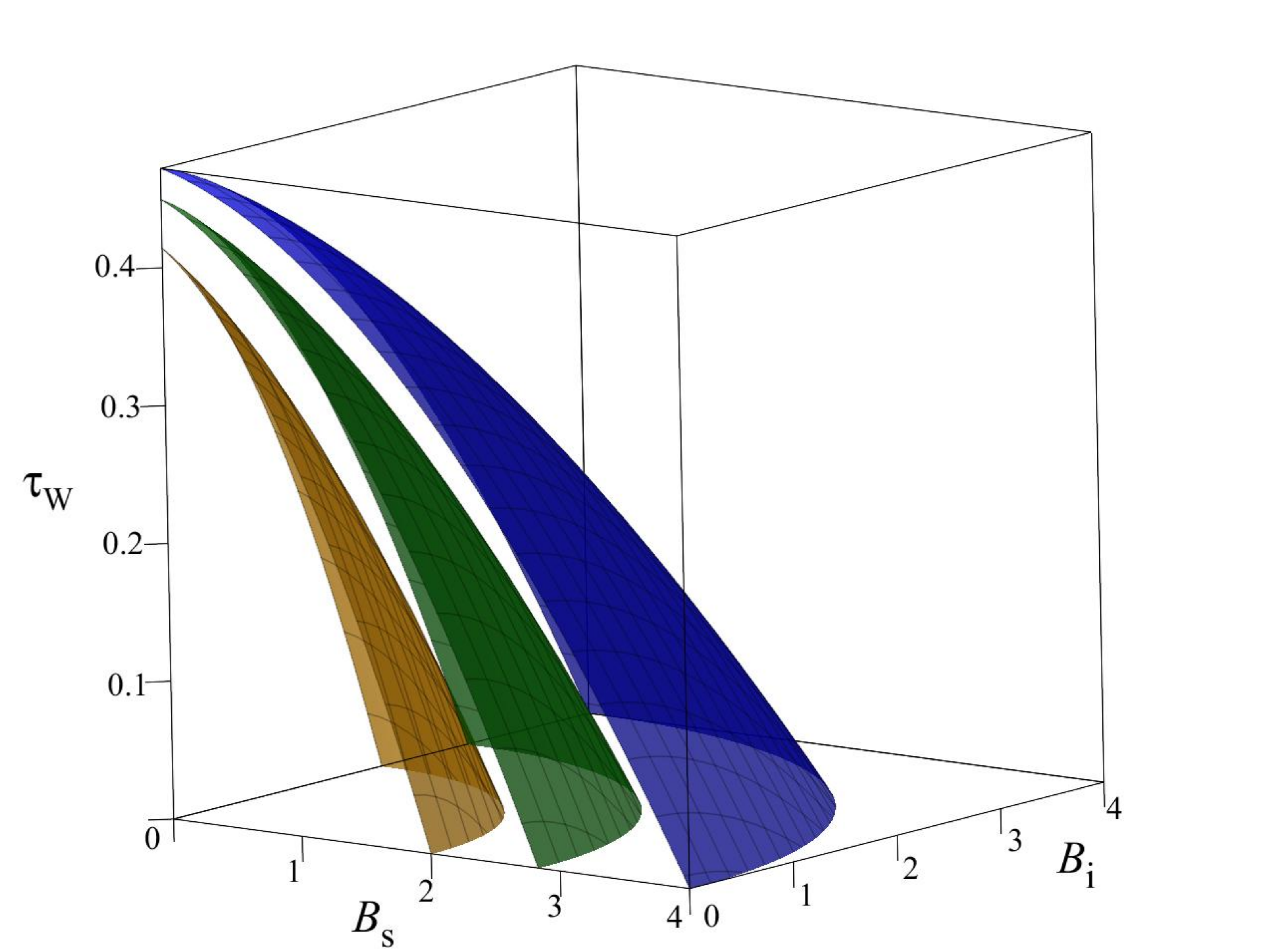}
 \caption{(Color online) Nonclassical intensity depth $\tau_W$ as a function of the mean noise
  photon numbers $B_{\rm s}$ and $B_{\rm i}$ for the mean photon-pair
  number $B_{\rm p}$ equal to $2$ [bottom, light-gray (yellow) area], $4$ [gray (green) area], and $8$
  [top, dark-gray (blue) area], assuming $M_{\rm p}=M_{\rm s}=M_{\rm i} =1$. The greater the
  value of $B_{\rm p} $, the greater the value of $ \tau_W$.}
\label{fig11}
\end{figure}

The nonclassical intensity depth $\tau_W $ has been determined for
the experimental multimode twin beams in
Ref.~\cite{PerinaJr2013a},
\begin{eqnarray}                
 \tau_W &=& \sqrt{\beta^2-\gamma}-\beta ,
\label{39}
\end{eqnarray}
where
\begin{eqnarray}                
 & & \beta = \frac{M_{\rm s}B_{\rm s}+M_{\rm i}B_{\rm i}+2M_{\rm p}B_{\rm p}}{M_{\rm s}+M_{\rm i}+2M_{\rm p}}, \nonumber \\
 & & \gamma =
  \frac{M_{\rm s}B_{\rm s}^{2}+M_{\rm i}B_{\rm i}^{2}-2M_{\rm p}B_{\rm p}}{M_{\rm s}+M_{\rm i}+2M_{\rm p}}.
\end{eqnarray}
The analysis of Eq.~(\ref{39}) shows that the experimental
multimode twin beam is strongly nonclassical ($ \tau_W > 0 $)
provided that
\begin{equation}  
 M_{\rm s} B_{\rm s}^2  + M_{\rm i} B_{\rm i}^2 < 2M_{\rm p} B_{\rm p} .
\label{41}
\end{equation}
Inequality (\ref{41}) means that the multimode strong
nonclassicality of the twin beam is lost if the noise is sufficiently
strong. For example, if $ M_{\rm p} = M_{\rm s} = M_{\rm i} $,
strongly nonclassical multimode twin beams are observed for $
B_{\rm s}^2 + B_{\rm i}^2 < 2B_{\rm p} $ (see Fig.~\ref{fig11}).
This behavior is similar to that discussed in Sec.~IV, though the
boundary given by $ \tau_W = 0 $ is quantitatively different
(compare Figs.~\ref{fig6} and \ref{fig11}). We also have here that
the greater the value of the mean photon-pair number $ B_{\rm p} $, the
greater the value of the nonclassical intensity depth $ \tau_W $.
Also, the greater the values of mean noise photon numbers $ B_{\rm
s} $ and $ B_{\rm i} $, the smaller the value of the nonclassical
intensity depth $ \tau_W $.

Similarly as in Sec.~VI, the logarithmic participation ratio $
R^{\rm log} $ can be defined for each component of the twin beam
to quantify its dimensionality. The logarithmic participation
ratio $ R^{\rm log} $ of the whole twin beam is then naturally
given as the sum of the logarithmic participation ratio $ R^{\rm
log}_{M_{\rm p},{\rm p}} $ of the paired component and the
logarithmic participation ratio $ R^{\rm log}_{M_{\rm s},{\rm s}}
+ R^{\rm log}_{M_{\rm i},{\rm i}} $ of the noise signal and idler
components. We note that Eq.~(\ref{Kfin}) is appropriate for
determining the participation ratio of both the single-mode noise signal
(or idler) field and the single-mode paired field. Alternatively we
may consider entropies of the components instead of participation
ratios. Entropies of the single-mode noise fields are given by
Eq.~(\ref{S}). Equation~(\ref{S}) is applicable also for
determination of the entropy of entanglement of a single-mode
paired field in a pure state for which $ \hat\rho_{{\rm s},jj}
\leftarrow c_j^2 $. As a consequence, the entropies $ S_{M_a,a} $ for
$a={\rm p,s,i}$, of each component are increasing functions of the
corresponding participation ratios $ R_{M_a,a} $. In single-mode
cases, these functions are determined by Eq.~(\ref{funSofK}),
plotted in Fig.~\ref{fig10}. Similarly to the overall logarithmic
participation ratio $ R^{\rm log} $, the overall entropy $ S $ can
be naturally split into its entangled part $S_{M_{\rm p},{\rm p}}$
and noisy part $ S_{M_{\rm s},{\rm s}} + S_{M_{\rm i},{\rm i}} $,
originating in the noise signal and idler components.

Finally, we briefly address the issue of the experimental
determination of the quantities discussed above. As these
quantities characterize the ``internal'' structure of a twin beam,
only their indirect determination is possible. It is based upon
the measurement of the joint signal-idler photocount histogram
using photon-number-resolving detectors. Knowing these detector
parameters~\cite{PerinaJr2012}, reconstruction of the joint
signal-idler photon-number
distribution~\cite{PerinaJr2012a,PerinaJr2013a} provides the
applied mean photon(-pair) numbers $ B $ and numbers $ M $ of
modes. The above-derived formulas then give the discussed
quantities.

\section{Conclusions}

The entanglement and nonclassicality of a single-mode noisy twin beam
have been quantified using the negativity and the nonclassical
depth, respectively. Universal mapping between the nonclassical
depth and the negativity has been revealed for noisy twin beams. The
mapping reflects the fact that nonclassicality of a twin beam is
caused by the entanglement of its two parts originating in pairing
of photons. Limitations to the amount of noise have been found to
preserve entanglement together with nonclassicality. the degrees of
freedom of a twin beam quantified by the signal- and idler-field
participation numbers have been divided into those needed to
describe entanglement and the remaining ones forming the noisy
signal and idler parts of the twin beam. The entanglement
dimensionality derived from the negativity has been applied here.
Entropy as an increasing function of the participation number has
been discussed. Properties of multimode twin beams have been
analyzed using appropriate quantities related per one mode. Also,
experimental multimode twin beams containing additional noise in
independent spatiotemporal modes have been investigated from the
point of view of their entanglement and multimode nonclassicality
including weak nonclassicality and dimensionality.

\acknowledgments I.A. and J.P. Jr. thank M. Bondani for discussions.
Support by project LO1305 of the Ministry of Education, Youth
and Sports of the Czech Republic is acknowledged. I.A. thanks
project IGA\_PrF\_2015004 of IGA UP Olomouc. J.P. Jr. acknowledges
project P205/12/0382 of GA \v{C}R. A.M. was supported by the Polish
National Science Centre under Grants No. DEC-2011/03/B/ST2/01903 and
No. DEC-2011/02/A/ST2/00305.

\end{document}